 \documentclass[12pt,draftclsnofoot,onecolumn]{IEEEtran}

\usepackage{amsfonts}
\usepackage{cite}
\usepackage{graphicx}
\usepackage{amssymb}
\usepackage{subfigure}
\usepackage{array}
\usepackage{color}
\usepackage{amsmath}
\usepackage{algorithmic}
\usepackage{algorithm}
\usepackage{cases}
\usepackage{bm}
\usepackage{mathrsfs}
\usepackage{epsfig}
\usepackage{psfrag}
\usepackage{url}
\usepackage{setspace}

\usepackage{etoolbox}
\makeatletter
\patchcmd{\@makecaption}
  {\scshape}
  {}
  {}
  {}
\makeatletter
\patchcmd{\@makecaption}
  {\\}
  {.\ }
  {}
  {}
\makeatother

\newcommand{\eqla}{\stackrel{(a)}{=}}
\newcommand{\eqlb}{\stackrel{(b)}{=}}
\newcommand{\eqlc}{\stackrel{(c)}{=}}
\newcommand{\eqld}{\stackrel{(d)}{=}}
\newcommand{\eqle}{\stackrel{(e)}{=}}
\newcommand{\eqlf}{\stackrel{(f)}{=}}
\newcommand{\eqlg}{\stackrel{(g)}{=}}
\newcommand{\eqlh}{\stackrel{(h)}{=}}
\newcommand{\eqli}{\stackrel{(i)}{=}}
\newcommand{\eqlj}{\stackrel{(j)}{=}}
\newcommand{\eqlk}{\stackrel{(k)}{=}}
\newcommand{\eqll}{\stackrel{(l)}{=}}

\newtheorem{Thm}{Theorem}

\newtheorem{Prob}{Problem}

\makeatletter
\renewcommand{\maketag@@@}[1]{\hbox{\m@th\normalsize\normalfont#1}}%
\makeatletter

\IEEEoverridecommandlockouts

\begin{document}

\title{Statistical Device Activity Detection for OFDM-based Massive Grant-Free Access}

\author{Yuhang Jia, {\em Student Member, IEEE}, Ying Cui, {\em Member, IEEE} and Wuyang Jiang
\footnote{Y. Jia and Y. Cui are with Shanghai Jiao Tong University, Shanghai, China.
W. Jiang is with Shanghai University of Engineering Science, Shanghai, China. The paper was presented in part at IEEE GLOBECOM 2021 \cite{YuhangGLOBECOM}.
}
}

\maketitle

 \begin{abstract}
Existing works on grant-free access, proposed to support massive machine-type communication (mMTC) for the Internet of things (IoT), mainly concentrate on narrow band systems under flat fading. However, little is known about massive grant-free access for wideband systems under frequency-selective fading. This paper investigates massive grant-free access in a wideband system under frequency-selective fading. First, we present an orthogonal frequency division multiplexing (OFDM)-based massive grant-free access scheme. Then, we propose two different but equivalent models for the received pilot signal, which are essential for designing various device activity detection and channel estimation methods for OFDM-based massive grant-free access. One directly models the received signal for actual devices, whereas the other can be interpreted as a signal model for virtual devices. Next, we investigate statistical device activity detection under frequency-selective Rayleigh fading based on the two signal models. We first model device activities as unknown deterministic quantities and propose three maximum likelihood (ML) estimation-based device activity detection methods with different detection accuracies and computation times. We also model device activities as random variables with a known joint distribution and propose three maximum a posterior probability (MAP) estimation-based device activity methods, which further enhance the accuracies of the corresponding ML estimation-based methods. Optimization techniques and matrix analysis are applied in designing and analyzing these methods. Finally, numerical results show that the proposed statistical device activity detection methods outperform existing state-of-the-art device activity detection methods under frequency-selective Rayleigh fading.
 \end{abstract}
 \begin{IEEEkeywords}
Massive machine-type communication (mMTC), massive grant-free access,
device activity detection, frequency-selective fading, orthogonal frequency division multiplexing (OFDM), maximum likelihood (ML) estimation, maximum a posterior probability (MAP) estimation
 \end{IEEEkeywords}





\section{Introduction}
With the rapid development of the Internet of Things (IoT), massive machine-type communication (mMTC) plays a vital role in the fifth generation (5G) cellular technologies and beyond. However, it is incredibly challenging to support enormous IoT devices which are energy-limited and sporadically active and have little data to send once activated. Grant-free access has been recently proposed to address the challenge. Specifically, devices are pre-assigned specific non-orthogonal pilots, active devices directly send their pilots, and the multi-antenna base station (BS) detects the active devices and estimates their channel conditions from the received pilot signal \cite{Erik}. Grant-free access successfully improves transmission efficiency and reduces device energy consumption.
Unfortunately, detecting colliding devices with non-orthogonal pilots at the BS is highly complicated.

Due to inherent sparse device activities in massive grant-free access, device activity detection and channel estimation can be formulated as compressed sensing (CS) problems and solved by CS-based algorithms, such as approximate message passing (AMP) \cite{Chen18TSP,Liu18TSP,Senel18TCOM,Shao19IoTJ,Chen19TWC} and   GROUP LASSO \cite{8536396,JSAC_li,9413733,9522070}. For instance, \cite{Chen18TSP,Liu18TSP,Senel18TCOM,Shao19IoTJ,Chen19TWC} propose AMP algorithms to deal with joint device activity detection and channel estimation in a single-cell network \cite{Chen18TSP,Liu18TSP,Senel18TCOM,Shao19IoTJ} and non-cooperative and cooperative device activity detection in a multi-cell network \cite{Chen19TWC}. In \cite{JSAC_li,9413733,8536396,9522070}, the authors apply GROUP LASSO to handle joint device activity detection and channel estimation \cite{JSAC_li,9413733,8536396} and  joint activity and data detection \cite{9522070} in a single-cell network.

Besides,  classic statistical estimation methods, such as maximum likelihood (ML) estimation \cite{Caire18ISIT,Yu19ICC,zhongGLOBECOM,Yu21SPAWC} and  maximum a posteriori probability (MAP) estimation \cite{JiangSPAWC,Jiang21TWC}, are applied for device activity detection.  It is noteworthy that device activity detection is a more fundamental path problem as channel conditions of the detected active devices can be subsequently estimated using conventional channel estimation methods. For example, \cite{Caire18ISIT} formulates device activity detection in a single-cell network as an ML estimation problem and proposes a computationally efficient iterative algorithm to obtain a stationary point of the ML estimation problem using the coordinate descent method.  In \cite{Yu19ICC}, the authors extend the ML estimation-based device activity detection method in \cite{Caire18ISIT} to joint activity and data detection in a single-cell network.
On the other hand,  in \cite{JiangSPAWC}, the authors formulate device activity detection with a general prior activity distribution in a single-cell network as an MAP estimation problem and extend the coordinate descent-based algorithm in \cite{Caire18ISIT} to obtain a stationary point of the MAP estimation problem. The proposed MAP estimation-based method in \cite{JiangSPAWC} can also be applied for joint activity and data detection in a single-cell network. Later, in  \cite{Jiang21TWC,zhongGLOBECOM,Yu21SPAWC}, the authors investigate non-cooperative \cite{Jiang21TWC} and cooperative \cite{Jiang21TWC,zhongGLOBECOM,Yu21SPAWC}
  device activity detection in multi-cell networks. Specifically, \cite{Jiang21TWC} proposes ML and MAP estimation-based methods to jointly estimate device activities and inter-cell or inter-cluster interference. In contrast, \cite{zhongGLOBECOM} and \cite{Yu21SPAWC} propose ML estimation-based methods to estimate device activities without considering inter-cluster interference. Notice that statistical estimation methods generally outperform CS-based estimation methods at the cost of computation complexity increase.

Recently, machine learning techniques have been  applied to resolve device activity detection and channel estimation in a single-cell network \cite{LiSPL,ZhangSPAWC,LiWCNC,JSAC_li,9484069,9508782,9146533,9432908}. For instance, data-driven \cite{LiSPL} and model-driven \cite{ZhangSPAWC,LiWCNC,JSAC_li} approaches are proposed to jointly design pilots and device activity detection and channel estimation methods using auto-encoder in deep learning. More specifically, the authors in \cite{ZhangSPAWC,LiWCNC,JSAC_li} propose multiple model-driven decoders, such as MAP-based, GROUP LASSO-based, and AMP-based decoders, for designing device activity detection and channel estimation methods. Both approaches can effectively utilize features of sparsity patterns embedded in data samples. Besides, the model-drive approach can significantly benefit from the underlying state-of-the-art methods, such as MAP, GROUP LASSO, and AMP. The joint design approaches can also be used for designing device activity detection and channel estimation methods for given pilots or designing pilots for given device activity detection and channel estimation methods. In later works \cite{9484069,9508782,9146533,9432908}, extended versions of the GROUP LASSO-based decoder \cite{9484069,9508782} and AMP-based decoder \cite{9146533,9432908} are also proposed for designing device activity detection and channel estimation methods for given pilots. Although generally achieving higher accuracies with shorter computation times, machine learning-based methods require time-consuming training over numerous data samples and have no theoretical performance guarantee.


Notice that most existing works \cite{Liu18TSP,Chen18TSP,Chen19TWC,Shao19IoTJ,Senel18TCOM,LiSPL,ZhangSPAWC,LiWCNC,JSAC_li,9413733,8536396,9522070,Caire18ISIT,Yu19ICC,zhongGLOBECOM,Yu21SPAWC,JiangSPAWC,Jiang21TWC,9146533,9432908,9484069,9508782} consider massive grant-free  access for a narrow band system under flat fading. However,
due to the signal corruption under frequency-selective fading, the existing methods for activity detection and channel estimation designed for a narrow band system under flat fading are no longer applicable for a wideband system under frequency-selective fading.  On the other hand, orthogonal frequency division multiplexing (OFDM) provides a high degree of robustness against channel-frequency selectivity. It hence is an attractive  choice for 4G-LTE and 5G-NR \cite{dahlman20205g}.
Very recently, \cite{9296241} and the early version of this work present OFDM-based grant-free access schemes for a wideband system under frequency-selective fading and propose ML estimation-based device activity detection methods for a single-cell network. Specifically, the authors in \cite{9296241} ignore the relationship among multiple paths for each device for tractability and solve the maximization of  an upper bound of the log-likelihood function under a specific frequency-selective fading model.  On the contrary, in \cite{YuhangGLOBECOM}, we precisely characterize the relationship among multiple channel taps for each device and maximize the exact log-likelihood function under frequency-selective Rayleigh fading. In this paper, we present a more extensive study on statistical device activity detection for OFDM-based massive grant-free access using ML estimation and MAP estimation. Specifically, we consider a single-cell network with one multi-antenna BS and $N$ single-antenna devices. We study the massive access scenario in a  wideband system under frequency-selective fading with $P$ channel taps.

{\bf OFDM-based Grant-free Access:}
 We present an OFDM-based massive grant-free access scheme and obtain two different but equivalent models for the received pilot signal under frequency-selective fading.  One directly models the received pilot signal from the $N$ actual devices, whereas the other can be interpreted as a received pilot signal model for $NP$ virtual devices experiencing flat fading. In particular, each actual device corresponds to $P$ virtual devices with an identical activity state. Based on the two signal models, we investigate statistical device activity detection under frequency-selective Rayleigh fading in this paper.  It is worth noting that based on the proposed signal model for virtual devices, we can directly apply standard channel estimation methods to estimate the channel conditions of the detected active devices and design CS-based device activity detection and channel estimation methods. This highlights the critical role of the proposed signal model for virtual devices in OFDM-based massive grant-free access.

    {\bf ML Estimation-based Device Activity Detection:}  We model  device activities as unknown deterministic quantities and propose three ML estimation-based device activity detection methods. Specifically, based on the two signal models for actual and virtual devices, we formulate two ML estimation problems for activity detections of actual and virtual devices, respectively. Then, we propose an iterative algorithm to obtain a stationary point of the non-convex ML estimation problem for activity detection of actual devices. We also propose two iterative algorithms of different computational complexities to obtain stationary points of an equivalent problem and an approximate problem of the non-convex ML estimation problem for virtual devices, respectively. Optimization techniques, such as the coordinate descent method and penalty method, and matrix analysis are applied in designing these algorithms. The three ML estimation-based methods have different detection accuracies and computation times and complement one another in the tradeoff between performance and complexity.

   {\bf MAP Estimation-based Device Activity Detection:}    We model device activities as random variables with a known joint distribution expressed with the multivariate Bernoulli (MVB) model \cite{JiangSPAWC,NIPS2011_4209} and propose three MAP estimation-based device activity methods. Specifically, based on the known joint distribution of the device activities and the two signal models for actual and virtual devices, we formulate two MAP estimation problems for activity detections of actual and virtual devices, respectively. Then, utilizing similar techniques and leveraging on the tractable form of the  MVB model, we develop three iterative algorithms to solve the two non-convex MAP estimation problems (to different extent). The resulting three MAP estimation-based  methods further enhance the corresponding ML estimation-based methods by exploiting the prior distribution of the device activities.

{\bf Numerical Results:} Numerical results show that the proposed statistical device activity detection methods outperform existing state-of-the-art device activity detection methods under frequency-selective Rayleigh fading.  Besides, numerical results demonstrate that the three proposed ML/MAP estimation-based methods achieve different detection accuracies and computation times at different values of the system parameters and hence can meet diverse practical needs for OFDM-based massive grant-free access.

\textbf{Notations:}
We represent vectors by boldface lowercase letters (e.g., $\mathbf{x}$), matrices by boldface uppercase letters (e.g., $\mathbf{X}$), scalar constants by non-boldface letters (e.g., $x$), and sets by calligraphic letters (e.g., $\mathcal{X}$). $x_i$ represents the $i$-th element of vector $\mathbf{x}$, $X_{i,j}$ represents the $(i,j)$-th element of matrix $\mathbf{X}$, $\mathbf{X}_{i,:}$ represents the $i$-th row of matrix $\mathbf{X}$, and $\mathbf{X}_{:,i}$ represents the $i$-th column of matrix $\mathbf{X}$. $\mathbf X_{:,1:k}$ represents the matrix consisting of the first $k$ columns of matrix $\mathbf{X}$. $\mathbf{X}^H$ and $\text{tr}\left(\mathbf{X}\right)$ denote the conjugate transpose and trace of matrix $\mathbf{X}$, respectively. $\mathcal{CN}(\mu,\sigma^2)$ denotes the complex Gaussian distribution with mean $\mu$ and variance $\sigma^2$. $\rm diag \left(\mathbf{x}\right)$ is a diagonal matrix with the entries of $\mathbf{x}$ on its main diagonal. $\lvert \cdot \rvert$ denotes the modulus of a complex number. $\mathbb{C}$, $\mathbb{R}_{+}$, and $\mathbb{R}_{++}$ denote the complex field, non-negative real field, and positive real field, respectively. $\otimes$ denotes the Kronecker product. $\mathbf{I}_L$ and $\mathbf{e}_n$ denote the $L\times L$ identity matrix and a unit vector whose $n$-th component is $1$, all others $0$. $\text{Pr}[x]$ denotes the probability of event $x$.
\vspace{-3mm}
\section{System Model}
We consider a single-cell cellular network with one $M$-antenna BS and $N$ single-antenna IoT devices. Let $\mathcal M\triangleq \{1,2,\cdots,M\}$ and $\mathcal N\triangleq \{1,2,\cdots,N\}$.
For all $n\in\mathcal{N}$, let $g_n>0$ denote the large-scale fading power of the channel between device $n$ and the BS. Small-scale fading follows the block fading model, i.e., small-scale fading coefficients remain constant within each coherence block and are independent and identically distributed (i.i.d.) over coherence blocks. We consider a wideband system under frequency-selective fading. Let $P$ denote the number of channel taps, and let $\mathcal P\triangleq \{1,2,\cdots,P\}$.
Denote $h_{n,m,p} \in \mathbb{C}$ as the $p$-th coefficient of the channel impulse response of the channel between device $n$ and the BS over antenna $m$, for all $n\in\mathcal{N},m\in\mathcal{M}$, $p\in\mathcal{P}$.

We study the massive access scenario arising from mMTC, where very few devices among a large number of potential devices are active and access the BS in each coherence block. For all $n\in\mathcal{N}$, let $\alpha_n\in\{0,1\}$ denote the activity state of device $n$, where $\alpha_n=1$ indicates that device $n$ is active and $\alpha_n=0$ otherwise. In the considered massive access scenario, $\sum_{n\in\mathcal{N}} \alpha_n \ll N$, i.e., $\boldsymbol{\alpha}\triangleq (\alpha_n)_{n\in\mathcal N} \in \{0,1\}^N$ is sparse. We adopt an OFDM-based massive grant-free  access scheme. Let $L$ denote the number of subcarriers, and denote $\mathcal L\triangleq \{1,2,\cdots,L\}$ as the set of subcarrier indices. Assume $P<L$. Each device $n\in\mathcal{N}$ is pre-assigned a specific pilot sequence $\tilde{\mathbf s}_n\triangleq(\tilde{s}_{n,\ell})_{\ell\in\mathcal L}\in \mathbb C^{L}$ consisting of $L\ll N$ OFDM symbols, each carried by one subcarrier. In the pilot transmission phase, active devices simultaneously send their length-$L$ pilots to the BS over the $L$ subcarriers, and the BS detects the activity  states of all devices and estimates the channel states of all active devices from the $LM$ received  OFDM symbols over the $L$ subcarriers and $M$ antennas.

The time domain representation of the OFDM symbols in $\tilde{\mathbf{s}}_n\in\mathbb{C}^L$, i.e., the normalized inverse discrete Fourier transform (IDFT) of $\tilde{\mathbf{s}}_n$, is given by:
\begin{align}
\mathbf s_n =\mathbf F^H\tilde{\mathbf s}_n \in \mathbb{C}^{L},\quad n \in \mathcal{N}.
\label{eq:IDFTzh}
\end{align}
Here, $\mathbf F \triangleq (F_{\ell,\ell'})_{\ell,\ell'\in\mathcal{L}} \in\mathbb C^{L\times L}$ denotes the discrete Fourier transform (DFT) matrix where $F_{\ell,\ell'}\triangleq \frac{1}{\sqrt{L}}e^{-\frac{j2\pi (\ell-1) (\ell'-1)}{L}}$. At each device $n\in\mathcal{N}$, a cyclic prefix is appended to $\mathbf s_n$ before transmission. After removing the signal corresponding to the cyclic prefixes,
the received signal over the $L$ signal dimensions at antenna $m\in\mathcal{M}$, $\mathbf r_m \triangleq (r_{\ell,m})_{\ell\in\mathcal L}\in\mathbb{C}^{L}$, can be written as \cite{8421267}:
\begin{align}
\mathbf r_m =& \sum_{n\in\mathcal N} \alpha_n g_n^{\frac{1}{2}}\mathbf H_{n,m}\mathbf s_n + \mathbf n_m = \sum_{n\in\mathcal N} \alpha_n g_n^{\frac{1}{2}}\mathbf H_{n,m}\mathbf{F}^{H} \tilde{\mathbf s}_n + \mathbf n_m, \quad m\in\mathcal{M},
\label{eq:receivesignal}
\end{align}
where
\begin{align}
\mathbf H_{n,m}\triangleq & \begin{bmatrix}
h_{n,m,1} & h_{n,m,L} &\cdots & h_{n,m,2}\\
h_{n,m,2} & h_{n,m,1} &\cdots & h_{n,m,3}\\
\vdots & \vdots &\ddots & \vdots\\
h_{n,m,L} & h_{n,m,L-1}&\cdots & h_{n,m,1}\\
\end{bmatrix} \in \mathbb{C}^{L\times L},\label{eq:matrixH}
\end{align}
and $\mathbf{n}_m \triangleq (n_{\ell,m})_{l\in\mathcal{L}} \in \mathbb{C}^{L}$ with $n_{\ell,m}\sim\mathcal {CN}(0,\sigma^2)$ is the  additive white Gaussian noise (AWGN).
Here, for notation convenience, we let $h_{n,m,p}=0$, $p \in \mathcal{L}\backslash\mathcal{P}, n\in\mathcal N$, $m\in\mathcal M$. Note that for all $n\in\mathcal{N},m\in\mathcal{M}$, each of $h_{n,m,l},l\in\mathcal{L}$ appears $L$ times in $\mathbf{H}_{n,m}$.

This paper aims to investigate device activity detection under frequency-selective fading, which is more challenging than device activity detection under flat fading \cite{Erik,Liu18TSP,JSAC_li,Caire18ISIT,Yu19ICC,Jiang21TWC,9146533,9432908}. Toward this end, based on $\mathbf{r}_m,m\in\mathcal{M}$ given in \eqref{eq:receivesignal}, this paper presents two different but equivalent models for the received pilot signal under frequency-selective fading which facilitate device activity detection. Specifically, one directly models the received pilot signal from the $N$ actual devices, whereas the other can be interpreted as a received pilot signal model for $NP$ virtual devices experiencing flat fading.

The signal model for the $N$ actual devices is presented as follows. Define $\tilde{\mathbf n}_m \triangleq \mathbf{F}\mathbf{n}_m\in\mathbb{C}^{L}$.
First, we obtain the received signal in the frequency domain, i.e.,
\begin{align}
\tilde{\mathbf r}_m=&\mathbf F\mathbf r_m = \sum_{n\in\mathcal N}\alpha_n g_n^{\frac{1}{2}}\mathbf F\mathbf H_{n,m}\mathbf F^H\tilde{\mathbf s}_n + \tilde{\mathbf n}_m
= \sum_{n\in\mathcal N}\alpha_ng_n^{\frac{1}{2}}{\rm diag}(\tilde{\mathbf{s}}_n)\mathbf{F}(\mathbf H_{n,m})_{:,1}+\tilde{\mathbf n}_m,\ m \in \mathcal{M}, \label{eqn:received_signal_frequency}
\end{align}
where the last equality is due to the fact that $\mathbf F\mathbf H_{n,m}\mathbf F^H \in \mathbb{C}^{L\times L}$ is a diagonal matrix \cite[Lemma 1]{8421267}. Define $\mathbf{S}_n \triangleq (\mathbf F^H{\rm diag}(\tilde{\mathbf{s}}_n)\mathbf F)_{:,1:P}$ and $\mathbf h_{n,m} \triangleq (h_{n,m,p})_{p\in\mathcal{P}}$. Then,
applying normalized IDFT to $\tilde{\mathbf r}_m$ in \eqref{eqn:received_signal_frequency}, we rewrite $\mathbf{r}_m$ in \eqref{eq:receivesignal} as:
\begin{align}
\mathbf r_m = & \mathbf F^H\tilde{\mathbf r}_m=\sum_{n\in\mathcal N} \alpha_n g_n^{\frac{1}{2}}\mathbf F^H{\rm diag}(\tilde{\mathbf{s}}_n)\mathbf F(\mathbf H_{n,m})_{:,1}+\mathbf n_m
 =  \sum_{n\in\mathcal N} \alpha_n g_n^{\frac{1}{2}} \mathbf{S}_n \mathbf h_{n,m} + \mathbf n_m, \ m\in\mathcal{M}, \label{eqn:received_signal_time}
\end{align}
where the last equality is due to $\mathbf{F}^H\mathbf{F}=\mathbf{I}_L$ and $h_{n,m,p}=0$, $p \in \mathcal{L}\backslash\mathcal{P}, n\in\mathcal N$, $m\in\mathcal M$.

The signal model for the $NP$ virtual devices is presented as follows. We construct $NP$ virtual devices.
Let $\mathcal{I} \triangleq \{1,...,NP\}$ denote the set of virtual devices. Let $\beta_i$ and $\delta_i$ denote the activity state and large-scale fading power of virtual device $i$, for all $i\in\mathcal{I}$. For all $n\in\mathcal{N}$, virtual devices $(n-1)P+1,...,nP$ share the same activity state and large-scale fading power as actual device $n$.
Thus, we have:
\begin{align}
& \beta_{(n-1)P+1}=...=\beta_{nP} = \alpha_n, \ n\in\mathcal{N},\label{eq:alpha_beta} \\
& \delta_{(n-1)P+1}=...=\delta_{nP} = g_n, \ n\in\mathcal{N}. \label{eq:virtual_g}
\end{align}
Besides, all virtual devices experience flat fading with the small-scale channel coefficient of virtual device $(n-1)P+p$ being $h_{n,m,p}$.
Therefore, the received signal from the $N$ devices, $\mathbf{r}_m$ in \eqref{eqn:received_signal_time}, can be equivalently rewritten as the received signal from the $NP$ virtual devices:
%
\begin{align}
\mathbf r_m
= & \mathbf S\mathbf B\mathbf{G}^{\frac{1}{2}}\mathbf h_m+\mathbf n_m, \quad m\in\mathcal{M}, \label{eqn:received_signal_time_xx}
\end{align}
where
$\mathbf{S}\triangleq \left[ \mathbf{S}_1,...,\mathbf{S}_N \right]$ $\in\mathbb{C}^{L\times NP}$,
$\mathbf B\triangleq {\rm diag}\left(\boldsymbol\beta\right)$ with $\boldsymbol\beta \triangleq (\beta_i)_{i\in\mathcal{I}}$, $\mathbf{G}\triangleq \rm{diag}\left(\mathbf{g}
\right)\otimes\mathbf I_P \in \mathbb{R}_{++}^{NP\times NP}$ with $\mathbf{g}\triangleq (g_n)_{n\in\mathcal{N}} \in \mathbb{R}_+^{N}$. By \eqref{eq:virtual_g}, we know $G_{i,i}=g_{\lceil\frac{i}{P}\rceil},i\in\mathcal{I}$.

Based on the two signal models for actual devices and virtual devices given by \eqref{eqn:received_signal_time} and \eqref{eqn:received_signal_time_xx}, respectively, this paper concentrates on statistical device activity detection. We consider a widely adopted frequency-selective fading model, namely frequency-selective Rayleigh fading, i.e., $h_{n,m,p}\sim \mathcal{CN}(0,1), n\in\mathcal{N}, m \in \mathcal{M}, p\in\mathcal{P}$, for tractability. We assume that the large-scale fading powers, $g_n,n\in\mathcal{N}$, are known to the BS for ease of exposition. The proposed methods can be readily extended to statistical device activity detection with unknown large-scale fading powers \cite{Caire18ISIT,Jiang21TWC}. Besides, we can further estimate the channel coefficients by estimating the channels of the detected virtual devices using MMSE. Last but not least, notice that we can also apply CS-based methods such as AMP and GROUP LASSO  (which has no requirement on the distributions of fading coefficients) for device activity detection and channel estimation, based on the proposed signal model for virtual devices.
These highlight the critical role of the proposed signal model for virtual devices in OFDM-based massive grant-free access.
\begin{table}[t]
\caption{\scriptsize{Computational complexities of proposed solutions as $N,L\rightarrow\infty$.}}\label{tab:computation_complexity}
\begin{center}
\vspace{-4mm}
\begin{scriptsize}
\textcolor{black}{\begin{tabular}{c|c}
\hline
Estimation Method & Computational Complexity \rule{0pt}{3mm}\\
\hline
ML Estimation-based Device Activity Detection for Actual Devices   & $\mathcal{O}(NPL^2)$\\
\hline
ML Estimation-based Device Activity Detection for Virtual Devices via Penalty Method & $\mathcal{O}(NPL^2)$ \\
\hline
ML Estimation-based Device Activity Detection for Virtual Devices via Relaxation Method & $\mathcal{O}(NPL^2$)\\
\hline
MAP Estimation-based Device Activity Detection for Actual Devices  & $\mathcal{O}(NPL^2+N2^N)$ \\
\hline
MAP Estimation-based Device Activity Detection for Virtual Devices via Penalty Method  & $\mathcal{O}(NPL^2+NP2^N)$ \\
\hline
MAP Estimation-based Device Activity Detection for Virtual Devices via Relaxation Method & $\mathcal{O}(NPL^2+NP2^N)$\\
\hline
\end{tabular}}
\end{scriptsize}
\end{center}
\vspace{-12mm}
\end{table}
\section{ML Estimation-based Device Activity Detection}\label{sec:ML}
In this section, we model $\alpha_n,n\in\mathcal{N}$ as unknown deterministic quantities and consider ML estimation for device activities. Specifically, based on the two signal models for actual devices and virtual devices, we propose three ML estimation-based device activity detection methods with computational complexities given in Table~\ref{tab:computation_complexity}.  Later in Section~\ref{sec:simulation}, we shall see that the three methods have different detection accuracies and computation times and complement one another in the tradeoff between performance and complexity.
\vspace{-3mm}
\subsection{ML Estimation-based Device Activity Detection for Actual Devices}\label{sec:non_extended}
In this part, based on the signal model for actual devices in \eqref{eqn:received_signal_time}, we propose an ML estimation-based device activity detection method that directly detects the activities of the $N$ actual devices, $\boldsymbol\alpha$. First, we formulate an ML estimation problem for device activity detection based on the expression of $\mathbf{r}_m$ in \eqref{eqn:received_signal_time}. Noting that
$\mathbf h_{n,m}, n\in\mathcal{N}, m\in\mathcal{M}$ are i.i.d. according to ${\mathcal {CN}}(\mathbf 0,\mathbf I_{P})$, $\mathbf r_m, m \in\mathcal{M}$, with $\mathbf{r}_m$ given by \eqref{eqn:received_signal_time}, are i.i.d. according to
$ \mathcal{CN}\left(\mathbf 0,\boldsymbol\Sigma^{(1)}_{\boldsymbol\alpha} \right)$ \cite{Caire18ISIT}, where
\begin{align}
\boldsymbol\Sigma^{(1)}_{\boldsymbol\alpha} \triangleq \sum_{n\in\mathcal N} \alpha_n g_n \mathbf{S}_n\mathbf S_n^H+\sigma^2\mathbf I_L. \label{eq:sigma_alpha_1}
\end{align}
Note that $\boldsymbol\Sigma^{(1)}_{\boldsymbol\alpha}$ depends on $\boldsymbol\alpha$. Let $\mathbf{R}$ with $\mathbf R_{:,m}\triangleq\mathbf{r}_m, m\in\mathcal{M}$ denote the received signal over the $L$ signal dimensions and $M$ antennas.
Thus, the likelihood function of $\mathbf  R$, viewed as a function of $\boldsymbol\alpha$, 
is given by:
\begin{align}
& p^{(1)}(\mathbf R;\boldsymbol\alpha) \triangleq \frac{\exp\left(-\text{tr}\left(
 \boldsymbol\Sigma^{(1)-1}_{\boldsymbol\alpha}\mathbf R\mathbf R^H\right)\right)}{\pi^{LM}\vert\boldsymbol\Sigma^{(1)}_{\boldsymbol\alpha}\vert^{M}}.\label{eqn:likelihood_no_coop}
\end{align}
The maximization of $p^{(1)}(\mathbf R;\boldsymbol\alpha)$ is equivalent to the minimization of $f^{(1)}(\boldsymbol\alpha)$, where
\begin{align}
f^{(1)}(\boldsymbol\alpha) \triangleq  -\log p^{(1)}(\mathbf R;\boldsymbol\alpha)-L\log\pi
=  \log|\boldsymbol\Sigma^{(1)}_{\boldsymbol\alpha}|+\text{tr}\left(\boldsymbol\Sigma_{\boldsymbol\alpha}^{(1)-1}\widehat{\mathbf \Sigma}_{\mathbf R}\right).\label{eqn:f_ml}
\end{align}
Here, $\widehat{\mathbf \Sigma}_{\mathbf R}\triangleq \frac{1}{M}\mathbf{R}\mathbf{R}^{H}$ represents the sample covariance matrix of $\mathbf{r}_m,m\in\mathcal{M}$. Note that $\widehat{\mathbf \Sigma}_{\mathbf R}$ is a sufficient statistics, since $f^{(1)}(\boldsymbol\alpha)$ depends on $\mathbf{R}$ only through $\widehat{\mathbf \Sigma}_{\mathbf R}$.
Thus, the ML estimation problem of $\boldsymbol\alpha$ can be formulated as follows.\footnote{In this paper, binary condition $\alpha_n\in\{0,1\}$ is relaxed to continuous condition $\alpha_n\in[0,1]$ in each estimation problem, and binary detection results are obtained by performing thresholding after solving the estimation problem as in~\cite{Caire18ISIT,Yu19ICC,JiangSPAWC,Jiang21TWC}.}
\begin{Prob}[ML Estimation for Activity Detection of Actual Devices]\label{Prob:ML_a_non_extension}
\begin{align}
\min_{\boldsymbol\alpha} &\quad f^{(1)}(\boldsymbol\alpha)\notag \\
s.t. &\quad   \alpha_{n} \in [0,1],\quad n \in\mathcal{N}. \label{eqn:a_n}
\end{align}
\end{Prob}

Problem~\ref{Prob:ML_a_non_extension} is a non-convex optimization problem over a convex set. When $P=1$, Problem~\ref{Prob:ML_a_non_extension} is equivalent to the ML estimation problem for activity detection of $N$ devices under flat Rayleigh fading in \cite{Caire18ISIT} and can be converted to the same form as the one in \cite{Caire18ISIT}. When $P\in\{2,3,...\}$, Problem~\ref{Prob:ML_a_non_extension} is different from the one in \cite{Caire18ISIT} and cannot be converted to its form (as $\mathbf{S}_n \mathbf{S}_n^H \in\mathbb{C}^{L\times L}$ is not a rank-one matrix).  Later, we shall see that this slight difference causes a significant challenge for solving Problem~\ref{Prob:ML_a_non_extension}.

Next, we adopt the coordinate descent method
to obtain a stationary point of Problem~\ref{Prob:ML_a_non_extension}.\footnote{The goal of solving a non-convex problem over a convex set is usually to obtain a stationary point of the problem.}
Specifically, given $\boldsymbol\alpha$ obtained in the previous step, the coordinate optimization with respect to $\alpha_n$ is equivalent to the optimization of the increment $d$ in $\alpha_n$:
\begin{align}
\min_{d\in [-\alpha_n,1-\alpha_n]} \ f^{(1)}(\boldsymbol\alpha + d\mathbf e_n).\label{eqn:Penal_a_ML_extension}
\end{align}
We shall see that it is more challenging to solve the coordinate optimization in~\eqref{eqn:Penal_a_ML_extension} for $P\in\{2,3,...\}$ than to solve that for $P=1$. In the following, we define two basic functions based on which we can characterize the optimal solution of the problem in \eqref{eqn:Penal_a_ML_extension}. Specifically, we first define:
\begin{align}
  f^{(1)}_{\boldsymbol\alpha,n}(d) \triangleq &
 \log |\mathbf{I}_P+dg_n\mathbf S_{n}^H\mathbf \Sigma^{(1)-1}_{\boldsymbol\alpha}\mathbf S_{n}| + dg_n\text{tr}\left((\mathbf{I}_P+dg_n\mathbf{S}_n^H\boldsymbol\Sigma_{\boldsymbol\alpha}^{(1)-1}
\mathbf{S}_n)^{-1}\mathbf{S}_n^H\boldsymbol\Sigma_{\boldsymbol\alpha}^{(1)-1}
\widehat{\mathbf \Sigma}_{\mathbf R}\boldsymbol\Sigma_{\boldsymbol\alpha}^{(1)-1}\mathbf{S}_n\right).\label{eq:function}
\end{align}
Applying eigenvalue decomposition, we can write:
\begin{align}
\mathbf{S}_n^H\boldsymbol\Sigma_{\boldsymbol\alpha}^{(1)-1}
\mathbf{S}_n=\mathbf{U}_n{\rm diag}(\mathbf{v})\mathbf{U}_n^H, \label{eq:eigenvalue_decomposition}
\end{align}
where $\mathbf{v} \triangleq (v_p)_{p\in\mathcal{P}} \in \mathbb{R}^{P}$ represents the eigenvalues, and $\mathbf{U}_n\in \mathbb{C}^{P\times P}$ represents the corresponding eigenvectors. For all $p\in\mathcal{P}$,
let $u_p$ denote the $p$-th diagonal element of $\mathbf{U}_n\mathbf{S}_n^H\boldsymbol\Sigma_{\boldsymbol\alpha}^{(1)-1}
\widehat{\mathbf \Sigma}_{\mathbf R}\boldsymbol\Sigma_{\boldsymbol\alpha}^{(1)-1}\mathbf{S}_n\mathbf{U}_n^H$.
Define $\mathbf{v}_{-p} \triangleq (v_{p'})_{p'\in\mathcal{P},p'\neq p} \in\mathbb{R}^{P-1}$,
\begin{align*}
\mathcal S(t) \triangleq & \left\{(\mathbf{x},\mathbf{y})|   \mathbf{x},\mathbf{y} \in \{0,1\}^{P-1}, x_p+y_p\leq 1,  p\in\mathcal{P}\backslash \{P\}, \sum_{p\in\mathcal{P}\backslash \{P\}}(x_p+2y_p)=t \right\},  \\
h(\mathbf{z},t)\triangleq & \sum_{(\mathbf{x},\mathbf{y})\in \mathcal S(t)}\prod_{p=1}^{P-1}2^{x_p}z_p^{x_p+2y_p},\ \mathbf{z} \in \mathbb{R}^{P-1},
\end{align*}
where $t=0,...,2P-2$.
Based on the above definitions, we then define:
\begin{align}
g^{(1)}_{\boldsymbol\alpha,n}(d) \triangleq &
d^{2P-1} \sum_{p\in\mathcal{P}}v_p^2h(\mathbf{v}_{-p},2P-1) +
\sum_{t=0}^{2P-2}d^t\sum_{p\in\mathcal{P}}
(v_p^2+v_p-u_p)h(\mathbf{v}_{-p},t).\label{eq:derivative_function}
\end{align}
Note that $g^{(1)}_{\boldsymbol\alpha,n}(d)$ is the numerator of the derivative of $f^{(1)}_{\boldsymbol\alpha,n}(d)$ (which is a fraction, as shown in Appendix A).
We write $f^{(1)}_{\boldsymbol\alpha,n}(d)$ and $g^{(1)}_{\boldsymbol\alpha,n}(d)$ as functions of $\boldsymbol\alpha$, as $\boldsymbol\Sigma_{\boldsymbol\alpha}^{(1)-1}$ is a
function of $\boldsymbol\alpha$. The optimal solution of the problem in~\eqref{eqn:Penal_a_ML_extension} is summarized below.
    \begin{Thm}[Optimal Solution of Coordinate Optimization in~\eqref{eqn:Penal_a_ML_extension}]\label{Thm:Step_APs_Penal_M_non_extension}
Given $\boldsymbol\alpha$ in the previous step, the optimal solution of the problem in~\eqref{eqn:Penal_a_ML_extension} is given by:
\begin{align}
d_n^{(1)} \triangleq \arg\min\limits_{d\in\mathcal{D}^{(1)}_n\cup \{-\alpha_n,1-\alpha_n\}} f^{(1)}_{\boldsymbol\alpha,n}(d),\label{eqn:d_Penalty_a_extension}
\end{align}
where $\mathcal{D}^{(1)}_n\triangleq \{d\in[-\alpha_n,1-\alpha_n]:g^{(1)}_{\boldsymbol\alpha,n}(d) = 0\}$.
\end{Thm}
\begin{IEEEproof}
Please refer to Appendix A.
\end{IEEEproof}

$g^{(1)}_{\boldsymbol\alpha,n}(d)$ is a polynomial with degree $2P-1$ and hence has $2P-1$ roots. Note that the roots of a polynomial with degree $q$ can be obtained analytically if $q\in\{1,2,3,4\}$ and numerically otherwise \cite{press2007numerical}. Besides, note that the computational complexities for obtaining the roots of a polynomial with degree $q$ analytically and numerically are $\mathcal{O}(q)$ and $\mathcal{O}(q^3)$, respectively \cite{press2007numerical}.  Thus, $\mathcal{D}^{(1)}_n$ can be obtained analytically with computational complexity $\mathcal{O}(P)$ if $P\in\{1,2\}$ and numerically with computational complexity $\mathcal{O}(P^3)$ otherwise.
The details of the coordinate descent method are summarized in Algorithm~\ref{alg:ML_descend_non_extension}.
\begin{algorithm}[t] \caption{\small{Algorithm for Solving Problem~\ref{Prob:ML_a_non_extension}}}
\scriptsize
\hspace*{0.02in} {\bf Input:}
empirical covariance matrix $\widehat{\mathbf \Sigma}_{\mathbf R}$. \\
\hspace*{0.02in} {\bf Output:}
activities of actual devices $\boldsymbol\alpha$.
\begin{algorithmic}[1]
\STATE Initialize $\mathbf \Sigma^{(1)-1}_{\boldsymbol\alpha}=\frac{1}{\sigma^2} \mathbf I_L$, $\boldsymbol\alpha=\mathbf 0$.
\STATE \textbf{repeat}
\FOR {$n\in\mathcal{N}$}
\STATE Calculate $d^{(1)}_n$ according to \eqref{eqn:d_Penalty_a_extension} analytically if $P\leq 2$ and numerically if $P\geq 3$.
\STATE \textbf{If} $d^{(1)}_n \neq 0$
\STATE \quad Update $\alpha_{n}=\alpha_{n}+d^{(1)}_n$.
\STATE \quad Update $\boldsymbol\Sigma^{(1)-1}_{\boldsymbol\alpha} = \boldsymbol \Sigma^{(1)-1}_{\boldsymbol\alpha}-d^{(1)}_ng_n
\boldsymbol \Sigma_{\boldsymbol\alpha}^{(1)-1}\mathbf S_{n}
(\mathbf{I}_P+ d^{(1)}_ng_n\mathbf S_{n}^H\boldsymbol\Sigma^{(1)-1}_{\boldsymbol\alpha}\mathbf S_{n})^{-1}\mathbf S_{n}^H\boldsymbol\Sigma^{(1)-1}_{\boldsymbol\alpha}$.
\STATE \textbf{end}
\ENDFOR
\STATE \textbf{until} $\boldsymbol\alpha$ satisfies some stopping criterion.
\end{algorithmic}\label{alg:ML_descend_non_extension}
\end{algorithm}
If each problem in \eqref{eqn:Penal_a_ML_extension} has a unique optimal solution, Algorithm~\ref{alg:ML_descend_non_extension} converges to a stationary point of Problem~\ref{Prob:ML_a_non_extension}, as the number of iterations goes to infinity \cite[Proposition 2.7.1]{Bertsekas99}.
The complexities of Step 4, Step 6, and Step 7 are $\mathcal{O}(PL^2)$, $\mathcal{O}(1)$, and $\mathcal{O}(PL^2)$, respectively, as $N,L\rightarrow\infty$.
Thus, the computational complexity of each iteration of Algorithm~\ref{alg:ML_descend_non_extension} is $\mathcal{O}(NPL^2)$.
\vspace{-3mm}
\subsection{ML Estimation-based Device Activity Detection for Virtual Devices}\label{sec:map_virtual}
In this part, based on the signal model for virtual devices, in \eqref{eqn:received_signal_time_xx}, we propose two ML estimation-based device activity detection methods which detect the activities of the $N$ actual devices, $\boldsymbol\alpha$, by detecting the activities of the $NP$ virtual devices, $\boldsymbol\beta$. First, we formulate an ML estimation problem for activity detection of the $NP$ virtual devices.
Noting that $\mathbf h_m \triangleq \left[\mathbf h^T_{1,m},...,\mathbf h^T_{N,m}\right]^T \in \mathbb{C}^{NP}$,
$\mathbf h_m$, $m\in\mathcal M$ are i.i.d. according to ${\mathcal {CN}}(\mathbf 0,\mathbf I_{NP})$, $\mathbf r_m, m \in\mathcal{M}$, with $\mathbf{r}_m$ given by \eqref{eqn:received_signal_time_xx}, are i.i.d. according to $ \mathcal{CN}\left(\mathbf 0,\boldsymbol\Sigma^{(2)}_{\boldsymbol\beta} \right)$ \cite{Caire18ISIT}, where
\begin{align}
\boldsymbol\Sigma^{(2)}_{\boldsymbol\beta} \triangleq \mathbf S\mathbf B\mathbf{G}\mathbf S^H+\sigma^2\mathbf I_L. \label{eq:sigma_alpha_2}
\end{align}
Thus, the likelihood function of $\mathbf R$, viewed as a function of $\boldsymbol\beta$, can be expressed as:
\begin{align}
&p^{(2)}(\mathbf R;\boldsymbol\beta) \triangleq \frac{\exp\left(-\text{tr}\left(\boldsymbol\Sigma_{\boldsymbol\beta}^{(2)-1}\mathbf R\mathbf R^{H}\right)\right)}{\pi^{LM}\vert\boldsymbol\Sigma
^{(2)}_{\boldsymbol\beta}\vert^M}.\label{eqn:likelihood_no_coop_xx}
\end{align}
The maximization of $p^{(2)}(\mathbf R;\boldsymbol\beta)$ is equivalent to the minimization of $f^{(2)}(\boldsymbol\beta)$, where
\begin{align}
 f^{(2)}(\boldsymbol\beta) \triangleq & -\log p^{(2)}(\mathbf R;\boldsymbol\beta) -L\log\pi
=   \log|\boldsymbol\Sigma^{(2)}_{\boldsymbol\beta}|+\text{tr}\left(\boldsymbol\Sigma_{\boldsymbol\beta}^{(2)-1}\widehat{\mathbf \Sigma}_{\mathbf R}\right). \label{eqn:f_ml_ex}
\end{align}Thus, the ML estimation problem of $\boldsymbol\beta$  from $\mathbf{R}$ given by \eqref{eqn:received_signal_time_xx} can be formulated as follows.
\begin{Prob}[ML Estimation for Activity Detection of Virtual Devices]\label{Prob:ML_a}
\begin{align}
\min_{\boldsymbol\beta} &\quad f^{(2)}(\boldsymbol\beta)\notag\\
s.t. &\quad  \beta_{(n-1)P+1}=...=\beta_{nP}, \  n\in\mathcal{N},\label{eq:equality_constraints} \\
& \quad  \beta_i \in [0,1],\quad  i\in\mathcal{I}. \label{eqn_beta_n}
\end{align}
\end{Prob}

Problem~\ref{Prob:ML_a} is also a non-convex optimization problem over a convex set. The objective function of Problem \ref{Prob:ML_a} differentiates from Problem~\ref{Prob:ML_a_non_extension}, as $\boldsymbol\Sigma^{(1)}_{\boldsymbol\alpha}$ and $\boldsymbol\Sigma^{(2)}_{\boldsymbol\beta}$ have different forms. Besides, the objective function of Problem \ref{Prob:ML_a} shares the same form as the objective function of the ML estimation problem for activity detection of $NP$ devices under  flat Rayleigh fading in \cite{Caire18ISIT} except that the dimensions of $\mathbf{S}\in\mathbb{C}^{L\times NP}$, $\mathbf{B}\in\mathbb{C}^{NP \times NP}$, $\mathbf{G}\in\mathbb{C}^{NP\times NP}$ rely on $P$. However, unlike Problem~\ref{Prob:ML_a_non_extension} and the ML estimation problem in \cite{Caire18ISIT}, Problem~\ref{Prob:ML_a} has extra coupling constraints in \eqref{eq:equality_constraints}.
%
%
Due to \eqref{eq:equality_constraints}, we cannot directly apply the coordinate descent method to solve Problem~\ref{Prob:ML_a}.
To address the issue caused by the coupling constraints in \eqref{eq:equality_constraints}, we apply the penalty method \cite{Bertsekas99} to obtain a stationary point of an equivalent problem of Problem~\ref{Prob:ML_a} in Section~\ref{subsec:penal} and adopt relaxation to obtain a stationary point of an approximate problem of Problem~\ref{Prob:ML_a} in Section~\ref{subsec:relax}, respectively. Later in Section~\ref{sec:simulation}, we shall see that the device activity detection method via the penalty method has higher accuracy and higher computational complexity than the method via relaxation. By \eqref{eq:alpha_beta}, we can construct the activities of the $N$ actual devices, $\boldsymbol\alpha$, according to:
\begin{align}
& \alpha_n = \frac{\sum_{p\in\mathcal{P}}\beta_{(n-1)P+p}}{P},\quad n\in\mathcal{N}, \label{eq:construct_alpha}
\end{align}
after solving Problem~\ref{Prob:ML_a} for $\boldsymbol\beta$.
\subsubsection{Activity Detection for Virtual Devices via Penalty Method}\label{subsec:penal}
First, we
disregard the coupling constraints in \eqref{eq:equality_constraints} and add a penalty for violating them to the objective function of Problem~\ref{Prob:ML_a}. Then, we can convert Problem~\ref{Prob:ML_a} to the following problem.
\begin{Prob}[Penalty Problem of Problem~\ref{Prob:ML_a}]\label{Prob:penal}
\begin{align*}
\min_{\mathbf b}\quad & \tilde{f}^{(2)}(\boldsymbol\beta)\triangleq f^{(2)}(\boldsymbol\beta)
+\rho \eta(\boldsymbol\beta)\notag \\
s.t. \quad &   \eqref{eqn_beta_n},
\end{align*}
where $\rho>0$ is the penalty parameter, and
\begin{align}
\eta(\boldsymbol\beta) \triangleq \sum_{n\in\mathcal{N}} \frac{\sum\limits_{p\in\mathcal{P}}\beta_{(n-1)P+p}}{P}
\bigg(1-\frac{\sum\limits_{p\in\mathcal{P}}\beta_{(n-1)P+p}}{P}\bigg) \label{eq:penaltyfunction}
\end{align}
is the penalty function.
\end{Prob}

If $\rho$ is sufficiently large, an optimal solution of Problem~\ref{Prob:penal} is also optimal for Problem~\ref{Prob:ML_a} (as $f^{(2)}(\boldsymbol\beta)$ is bounded from above) \cite{Bertsekas99}.
Now, we adopt the coordinate descent method to obtain a stationary point of Problem~\ref{Prob:penal} instead of Problem~\ref{Prob:ML_a}. Specifically, given $\boldsymbol\beta$ obtained in the previous step, the coordinate descent optimization with respect to $\beta_i$ is equivalent to the optimization of the increment $d$ in $\beta_i$:
\begin{align}
\min_{d\in [-\beta_i,1-\beta_i]} \ \tilde{f}^{(2)}(\boldsymbol\beta + d\mathbf e_i).\label{eqn:Penal_a}
\end{align}
We shall see that it is more challenging to solve the coordinate optimization in~\eqref{eqn:Penal_a}  than to solve the ML estimation problem for flat Rayleigh fading in \cite{Caire18ISIT}. Similarly, we define two important functions before solving the problem:
\begin{align}
\tilde{f}^{(2)}_{\boldsymbol\beta,i}(d) \triangleq &\log(1+d\delta_i\mathbf S_{:,i}^H\mathbf \Sigma^{(2)-1}_{\boldsymbol\beta}\mathbf S_{:,i}) -\frac{d\delta_i\mathbf S_{:,i}^H\mathbf \Sigma^{(2)-1}_{\boldsymbol\beta}\widehat{\mathbf \Sigma}_{\mathbf R}\mathbf \Sigma^{(2)-1}_{\boldsymbol\beta}\mathbf S_{:,i} }{1+d\delta_i\mathbf S_{:,i}^H\mathbf \Sigma^{(2)-1}_{\boldsymbol\beta}\mathbf S_{:,i}}+\frac{\rho d}{P}
\bigg(1-\frac{d}{P}-\frac{2}{P}\sum\limits_{p=1}^{P}\beta_{\left(\lceil \frac{i}{P} \rceil-1\right)P+p}\bigg), \label{eq:f_old}\\
  \tilde{g}^{(2)}_{\boldsymbol\beta,i}(d) \triangleq &
   A^{(2)}_id^3+B^{(2)}_id^2+C^{(2)}_id+D^{(2)}_i, \label{eq:three_times}
\end{align}
where
\begin{align}
 & A^{(2)}_i\triangleq -\frac{2\rho \delta_i^2}{P^2}\left(\mathbf S_{:,i}^H\mathbf \Sigma^{(2)-1}_{\boldsymbol\beta}\mathbf S_{:,i}\right)^2, \notag \\
& B^{(2)}_i\triangleq \frac{\rho \delta_i^2}{P}\bigg(1-\frac{2}{P}\sum\limits_{p=1}^{P}\beta_{\left(\lceil \frac{i}{P} \rceil-1\right)P+p}\bigg)\left(\mathbf S_{:,i}^H\mathbf \Sigma^{(2)-1}_{\boldsymbol\beta}\mathbf S_{:,i}\right)^2-\frac{4\rho \delta_i}{P^2}\mathbf S_{:,i}^H\mathbf \Sigma^{(2)-1}_{\boldsymbol\beta}\mathbf S_{:,i},\notag \\
&C^{(2)}_i\triangleq \delta_i^2 \left(\mathbf S_{:,i}^H\mathbf \Sigma^{(2)-1}_{\boldsymbol\beta}\mathbf S_{:,i}\right)^2+\frac{2\rho \delta_i}{P}\bigg(1-\frac{2}{P}\sum\limits_{p=1}^{P}\beta_{\left(\lceil \frac{i}{P} \rceil-1\right)P+p}\bigg)\mathbf S_{:,i}^H\mathbf \Sigma^{(2)-1}_{\boldsymbol\beta}\mathbf S_{:,i}-\frac{2\rho}{P^2},\notag \\
 &D^{(2)}_i\triangleq \delta_i\mathbf S_{:,i}^H\mathbf \Sigma^{(2)-1}_{\boldsymbol\beta}\mathbf S_{:,i} -\delta_i\mathbf S_{:,i}^H\mathbf \Sigma^{(2)-1}_{\boldsymbol\beta}\widehat{\mathbf \Sigma}_{\mathbf R}\mathbf \Sigma^{(2)-1}_{\boldsymbol\beta}\mathbf S_{:,i} +\frac{\rho}{P}\bigg(1-\frac{2}{P}\sum\limits_{p=1}^{P}\beta_{\left(\lceil \frac{i}{P} \rceil-1\right)P+p}\bigg).\notag
\end{align}
Note that $\tilde{g}^{(2)}_{\boldsymbol\beta,i}(d)$ is the numerator of the derivative of $\tilde{f}^{(2)}_{\boldsymbol\beta,i}(d)$ (which is a fraction, as shown in Appendix B). The optimal solution of the problem in~\eqref{eqn:Penal_a} is summarized below.
\begin{Thm}[Optimal Solution of Coordinate Optimization in~\eqref{eqn:Penal_a}]\label{Thm:Step_APs_Penal_M}
Given $\mathbf b$ in the previous step, the optimal solution of the problem in~\eqref{eqn:Penal_a} is given by:
\begin{align}
d^{(2)}_i = \arg\min\limits_{d\in\mathcal{D}^{(2)}_i\cup\{-\beta_i,1-\beta_i\}} \tilde{f}^{(2)}_{\boldsymbol\beta,i}(d),\label{eqn:d_Penalty_a}
\end{align}
where $\mathcal{D}^{(2)}_i \triangleq \{d\in[-\beta_i,1-\beta_i]:\tilde{g}^{(2)}_{\boldsymbol\beta,i}(d) = 0\}$.
\end{Thm}
\begin{IEEEproof}
Please refer to Appendix B.
\end{IEEEproof}

As $\tilde{g}^{(2)}_{\boldsymbol\beta,i}(d)$ is a polynomial with degree 3, $\mathcal{D}^{(2)}_i$ can be obtained in closed-form with computational complexity $\mathcal{O}(P)$.
The details of the coordinate descent method are summarized in Algorithm~\ref{alg:ML_descend}.
\begin{algorithm}[t] \caption{\small{Algorithm for Solving Problem~\ref{Prob:penal} and Problem~\ref{Prob ML_relax}}}
\scriptsize
\hspace*{0.02in} {\bf Input:}
empirical covariance matrix $\widehat{\mathbf \Sigma}_{\mathbf R}$. \\
\hspace*{0.02in} {\bf Output:}
activities of virtual devices $\boldsymbol\beta$.
\begin{algorithmic}[1]
\STATE Initialize $\mathbf \Sigma^{(2)-1}_{\boldsymbol\beta}=\frac{1}{\sigma^2} \mathbf I_L$, $\boldsymbol\beta=\mathbf 0$.
\STATE \textbf{repeat}
\FOR {$i\in\mathcal{I}$}
\STATE {\bf For Problem~\ref{Prob:penal}:} Calculate $d^{(2)}_i$ according to \eqref{eqn:d_Penalty_a}. \\
{\bf For Problem~\ref{Prob ML_relax}:} Calculate $d^{(2)}_i$ according to \eqref{eqn:d_ML_relax}.
\STATE \textbf{If} $d^{(2)}_i\neq 0$
\STATE \quad Update $\beta_{i}=\beta_{i}+d^{(2)}_i$.
\STATE \quad Update $\mathbf \Sigma^{(2)-1}_{\boldsymbol\beta} = \mathbf \Sigma^{(2)-1}_{\boldsymbol\beta}-\frac{d^{(2)*}_i\delta_i\mathbf \Sigma_{\boldsymbol\beta}^{(2)-1}\mathbf S_{:,i}\mathbf S_{:,i}^H\mathbf \Sigma^{(2)-1}_{\boldsymbol\beta}}{1+d^{(2)*}_i\delta_i\mathbf S_{:,i}^H\mathbf \Sigma^{(2)-1}_{\boldsymbol\beta}\mathbf S_{:,i}}$.
\STATE \textbf{end}
\ENDFOR
\STATE \textbf{until} $\boldsymbol\beta$ satisfies some stopping criterion.
\end{algorithmic}\label{alg:ML_descend}
\end{algorithm}
If each problem in \eqref{eqn:Penal_a} has a unique optimal solution, Algorithm~\ref{alg:ML_descend} for solving Problem~\ref{Prob:penal} converges to a stationary point of Problem~\ref{Prob:penal}, as the number of iterations goes to infinity \cite[Proposition 2.7.1]{Bertsekas99}.
The computational complexities of Step 4, Step 6, and Step 7 are $\mathcal{O}(L^2)$, $\mathcal{O}(1)$, and $\mathcal{O}(L^2)$, respectively, as $L\rightarrow\infty$. Thus, the computational complexity of each iteration of Algorithm \ref{alg:ML_descend} for solving Problem~\ref{Prob:penal} is $\mathcal{O}\left(NPL^2\right)$, as $N,L\rightarrow\infty$.
\vspace{-1mm}
\subsubsection{Activity Detection for Virtual Devices via Relaxation} \label{subsec:relax}
First, we relax the coupling constraints in \eqref{eq:equality_constraints} of Problem~\ref{Prob:ML_a} as in \cite{Yu19ICC}. The relaxed version of Problem~\ref{Prob:ML_a} is given as follows.
\begin{Prob}[Relaxed Problem of Problem~\ref{Prob:ML_a}]\label{Prob ML_relax}
\begin{align}
\min_{\boldsymbol\beta} &\quad f^{(2)}(\boldsymbol\beta)\notag\\
s.t.  & \quad \eqref{eqn_beta_n}.\notag
\end{align}
\end{Prob}

Obviously, Problem~\ref{Prob ML_relax} has the same  variables and constraints but a simpler objective function than Problem~\ref{Prob:penal}. Thus, we expect that Problem~\ref{Prob ML_relax} can be solved with lower computational complexity than  Problem~\ref{Prob:penal}. Actually, Problem~\ref{Prob ML_relax} shares the same form as the ML estimation problem for activity detection of $NP$ devices under flat Rayleigh fading in \cite{Caire18ISIT}. Thus, the coordinate descent method in \cite[Algorithm 1]{Caire18ISIT} can be directly used to obtain a stationary point of Problem~\ref{Prob ML_relax}. We present the coordinate descent method here for completeness. Specifically, given $\boldsymbol\beta$ obtained in the previous step, the coordinate optimization with respect to $\beta_i$ is equivalent to the optimization of the increment $d$ in $\beta_i$:
\begin{align}
\min_{d\in [-\beta_i,1-\beta_i]} \ f^{(2)}(\boldsymbol\beta + d\mathbf e_i).\label{eqn:Penal_a_relax}
\end{align}
Its optimal solution is given by \cite{Caire18ISIT}:
\begin{align}
d^{(2)}_i = \min\left\{\max\left\{\frac{\mathbf S_{:,i}^H\mathbf \Sigma^{(2)-1}_{\boldsymbol\beta}\widehat{\mathbf \Sigma}_{\mathbf R}\mathbf \Sigma^{(2)-1}_{\boldsymbol\beta}\mathbf S_{:,i}-\mathbf S_{:,i}^H\mathbf \Sigma^{(2)-1}_{\boldsymbol\beta}\mathbf S_{:,i}}{\delta_i(\mathbf S_{:,i}^H\mathbf \Sigma^{(2)-1}_{\boldsymbol\beta}\mathbf S_{:,i})^2},-\beta_{i}\right\},1-\beta_i\right\}.\label{eqn:d_ML_relax}
\end{align}

The details of the coordinate descent method for solving Problem~\ref{Prob ML_relax} are also summarized in Algorithm~\ref{alg:ML_descend}.
As $f^{(2)}(\boldsymbol\beta)$ is continuously differentiable, and the coordinate optimization in \eqref{eqn:Penal_a_relax} has a unique optimal solution, we know that Algorithm~\ref{alg:ML_descend} for solving Problem~\ref{Prob ML_relax} converges to a stationary point of Problem~\ref{Prob ML_relax}, as the number of iterations goes to infinity \cite[Proposition 2.7.1]{Bertsekas99}. The computational complexities of Step 4, Step 6, and Step 7 are $\mathcal{O}(L^2)$, $\mathcal{O}(1)$, and $\mathcal{O}(L^2)$, respectively, as $L\rightarrow\infty$. Thus, the computational complexity of each iteration of Algorithm \ref{alg:ML_descend} for solving Problem~\ref{Prob ML_relax} is $\mathcal{O}\left(NPL^2\right)$, as $N,L \rightarrow \infty$.
\addtolength{\topmargin}{0.025in}
\vspace{-3mm}
\section{MAP Estimation-based Device Activity Detection}\label{sec:map}
In this section, we model $\alpha_n,n\in\mathcal{N}$ as random variables with a known joint distribution and consider MAP estimation for device activities. Specifically, as in \cite{JiangSPAWC}, we adopt a general and tractable model for $p\left(\boldsymbol \alpha\right)$, i.e., the multivariate Bernoulli (MVB) model~\cite{NIPS2011_4209}. The probability mass function (p.m.f.) of $\boldsymbol\alpha$ under the MVB model is given by:
\begin{align}
p\left(\boldsymbol \alpha\right)=\exp\bigg(\sum_{\omega\in\Psi}\bigg(c_{\omega}\prod_{n\in\omega}\alpha_n\bigg)-s\bigg),
\label{eqn:pmf_a_typical_a}
\end{align}
where $\Psi$ denotes the set of nonempty subsets of $\mathcal{N}$,
$c_{\omega}$ is the coefficient reflecting the correlation among $\alpha_n$, $n\in\omega$, and
$s\triangleq \log(\sum_{\boldsymbol \alpha\in\{0,1\}^{N}}\exp(\sum_{\omega\in\Psi}(c_{\omega}\prod_{n\in\omega}\alpha_n)))$ is the normalization factor. Based on the MVB model in \eqref{eqn:pmf_a_typical_a}, we extend the three ML estimation-based device activity detection methods in Section~\ref{sec:ML} to MAP estimation-based device activity detection methods with computational complexities given in Table~\ref{tab:computation_complexity}. Similarly, the three MAP estimation-based methods have different detection accuracies and computation times and complement one another in the tradeoff between performance and complexity.
\vspace{-3mm}
\subsection{MAP Estimation-based Device Activity Detection for Actual Devices}\label{sec:map_actual}
In this part, based on the signal model for actual devices in \eqref{eqn:received_signal_time}, we propose an MAP estimation-based device activity detection method that directly detects the activities of the $N$ actual devices. First, we formulate an MAP estimation problem for device activity detection based on the expression of $\mathbf{r}_m$ in \eqref{eqn:received_signal_time}.
Based on  the conditional p.d.f. of $\mathbf{R}$ given $\boldsymbol\alpha$ (identical to the likelihood function of $\mathbf{R}$ given in \eqref{eqn:likelihood_no_coop}) and the prior distribution of $\boldsymbol\alpha$, the joint density of $(\mathbf{R},\boldsymbol\alpha)$ is given by \cite{JiangSPAWC}:
 \begin{align}
 &p^{(3)}(\mathbf R,\boldsymbol\alpha)
 \triangleq p^{(1)}(\mathbf R;\boldsymbol\alpha)p(\boldsymbol\alpha)
= \frac{\exp\left(-{\rm tr}\left(\boldsymbol\Sigma^{(1)-1}_{\boldsymbol\alpha}\mathbf R\mathbf R^H\right)+\sum_{\omega\in\Psi}c_{\omega}\prod_{n\in\omega}\alpha_n-s
 \right)}{\pi^{LM}\vert\boldsymbol\Sigma^{(1)-1}_{\boldsymbol\alpha}\vert^M}. \label{eqn:likelihood_no_coop_map}
 \end{align}
The maximization of $p^{(3)}(\mathbf R,\boldsymbol\alpha)$ is equivalent to the minimization of $f^{(3)}(\boldsymbol\alpha) $, where
\begin{align}\label{eqn:f_map}
  f^{(3)}(\boldsymbol\alpha) \triangleq  -\log p^{(3)}(\mathbf R,\boldsymbol\alpha)-L\log\pi = f^{(1)}(\boldsymbol\alpha)-\frac{1}{M}\sum_{\omega\in\Psi}c_{\omega}\prod_{n\in\omega}\alpha_n,
\end{align}
where $f^{(1)}(\boldsymbol\alpha)$ is given by \eqref{eqn:f_ml}.
Thus, the MAP estimation problem of $\boldsymbol\alpha$ from $\mathbf{r}_m$ given by \eqref{eqn:received_signal_time} can be formulated as follows.
\begin{Prob}[MAP Estimation  for Activity Detection of Actual Devices]\label{prob:map}
\begin{align*}
\min_{\boldsymbol\alpha} \quad &f^{(3)}(\boldsymbol\alpha) \\
s.t. \quad & \eqref{eqn:a_n}.
\end{align*}
\end{Prob}

Note that $f^{(3)}(\boldsymbol\alpha)-f^{(1)}(\boldsymbol\alpha)$ decreases with $M$. This indicates that the prior distribution of $\boldsymbol\alpha$ becomes less critical and the observations $\mathbf{R}$ become more critical, as $M$ increases. Thus,  the MAP estimation given by Problem~\ref{prob:map}  is effective when $M$ is not large, and $p(\boldsymbol\alpha)$ is available, compared to the ML estimation given by Problem~\ref{Prob:ML_a_non_extension}.
Problem~\ref{prob:map} is a non-convex optimization problem over a convex set.  When $P=1$, Problem~\ref{prob:map} is equivalent to the MAP estimation problem for activity detection of $N$ devices under flat Rayleigh fading in \cite{JiangSPAWC} and can be converted to the same form as the one in \cite{JiangSPAWC}. When $P\in\{3,4,...\}$, Problem~\ref{prob:map} is different from the one in \cite{JiangSPAWC} and cannot be converted to its form  (as $\mathbf{S}_n\mathbf{S}_n^H \in\mathbb{C}^{L\times L}$ is not a rank-one matrix).

Next, we adopt the coordinate descent method to obtain a stationary point of Problem~\ref{prob:map}. Specifically, given $\boldsymbol\alpha$ obtained in the previous step, the coordinate optimization with respect to $\alpha_n$ is equivalent to the optimization of the increment $d$ in $\alpha_n$:
\begin{align}\label{eqn:Penal_a_MAP_extension}
  \min_{d\in[-\alpha_n,1-\alpha_n]} f^{(3)}(\boldsymbol\alpha+d\mathbf{e}_n).
\end{align}
We shall see that it is more challenging to solve the coordinate optimization in \eqref{eqn:Penal_a_MAP_extension} for $P\in\{2,3,...\}$ than to solve that for $P=1$. To solve the
problem in \eqref{eqn:Penal_a_MAP_extension}, we first define:
\begin{align}
\epsilon_n(\boldsymbol\alpha) \triangleq & \sum_{\omega\in\Psi:n\in\omega}c_{\omega}\prod_{n^{'}\in\omega,n^{'}\neq n}\alpha_{n'}, \label{eq:mvb_coordinate} \\
  f^{(3)}_{\boldsymbol\alpha,n}(d) \triangleq &
f^{(1)}_{\boldsymbol\alpha,n}(d) - \frac{d\epsilon_n(\boldsymbol\alpha)}{M},  \label{eq:function_map}\\
  g^{(3)}_{\boldsymbol\alpha,n}(d) \triangleq & g^{(1)}_{\boldsymbol\alpha,n}(d) - \frac{\prod_{p\in\mathcal{P}}
  (1+v_pd)^2}{M}\epsilon_n(\boldsymbol\alpha),
  \label{eq:derivative_function_map}
\end{align}
where $f^{(1)}_{\boldsymbol\alpha,n}(d)$ and $g^{(3)}_{\boldsymbol\alpha,n}(d)$ are given by \eqref{eq:function} and \eqref{eq:derivative_function}, respectively. Note that $\epsilon_n(\boldsymbol\alpha)$ is  the derivative of $\sum_{\omega\in\Psi:n\in\omega}c_{\omega}(\alpha_n+d)\prod_{n^{'}\in\omega,n^{'}\neq n}\alpha_{n'}$,
and $g^{(3)}_{\boldsymbol\alpha,n}(d)$ is the numerator of the derivative of $f^{(3)}_{\boldsymbol\alpha,n}(d)$  (which is a fraction, as shown in Appendix C). The optimal solution of the problem in~\eqref{eqn:Penal_a_MAP_extension} is summarized below.
\begin{Thm}[Optimal Solution of Coordinate Optimization in~\eqref{eqn:Penal_a_MAP_extension}]\label{Thm:Step_APs_Penal_M_non_extension_map}
Given $\boldsymbol\alpha$, the optimal solution of the problem in~\eqref{eqn:Penal_a_MAP_extension} is given by:
\begin{align}
d_n^{(3)*} \triangleq \arg\min\limits_{d\in\mathcal{D}^{(3)}_n\cup \{-\alpha_n,1-\alpha_n\}} f^{(3)}_{\boldsymbol\alpha,n}(d),\label{eqn:d_Penalty_a_extension_map}
\end{align}
where $\mathcal{D}^{(3)}_n\triangleq \{d\in[-\alpha_n,1-\alpha_n]:g^{(3)}_{\boldsymbol\alpha,n}(d) = 0\}$.
\end{Thm}
\begin{IEEEproof}
Please refer to Appendix C.
\end{IEEEproof}

As $g^{(3)}_{\boldsymbol\alpha,n}(d)$ is a polynomial with degree $2P$, $\mathcal{D}^{(3)}_n$ can be obtained analytically with computational complexity $\mathcal{O}(P+2^N)$ if $P\in\{1,2\}$ and numerically with computational complexity $\mathcal{O}(P^3+2^N)$ if $P\in\{3,4,...\}$.
The details of the coordinate descent method are summarized in Algorithm~\ref{alg:MAP_descend_non_extension}. If each problem in \eqref{eqn:Penal_a_MAP_extension} has a unique optimal solution, Algorithm~\ref{alg:MAP_descend_non_extension} converges to a stationary point of Problem~\ref{prob:map}, as the number of iterations goes to infinity \cite[Proposition 2.7.1]{Bertsekas99}. The computational complexities of Step 4, Step 6, and Step 7 are $\mathcal{O}(PL^2+2^N)$, $\mathcal{O}(1)$, and $\mathcal{O}(PL^2+2^N)$, respectively, as $N,L\rightarrow\infty$.
Thus, the computational complexity of each iteration of Algorithm~\ref{alg:MAP_descend_non_extension}  is $\mathcal{O}(NPL^2+N2^N)$,  higher than that for solving Problem~\ref{Prob:ML_a_non_extension} (as the objective function incorporating the prior distribution of $\boldsymbol\alpha$ is more complex). Notice that the term $N2^N$ corresponds to the computational complexity for calculating $\epsilon_n(\boldsymbol\alpha)$ in \eqref{eq:mvb_coordinate}, which derives from the adopted MVB model in the most general form, i.e., $p(\boldsymbol\alpha)$ in \eqref{eqn:pmf_a_typical_a}. However, for some special cases of $p(\boldsymbol\alpha)$, the term $N2^N$ can be significantly reduced \cite{Jiang21TWC}. Besides, note that as $\boldsymbol\alpha$ is a sparse vector, the actual computational complexity for calculating $\epsilon_n(\boldsymbol\alpha)$ is much lower.
\begin{algorithm}[t] \caption{\small{Algorithm for Solving Problem~\ref{prob:map}}}
\scriptsize
\hspace*{0.02in} {\bf Input:}
empirical covariance matrix $\widehat{\mathbf \Sigma}_{\mathbf R}$. \\
\hspace*{0.02in} {\bf Output:}
activities of actual devices $\boldsymbol\alpha$.
\begin{algorithmic}[1]
\STATE Initialize $\mathbf \Sigma^{(1)-1}_{\boldsymbol\alpha}=\frac{1}{\sigma^2} \mathbf I_L$, $\boldsymbol\alpha=\mathbf 0$.
\STATE \textbf{repeat}
\FOR {$n\in\mathcal{N}$}
\STATE Calculate $d^{(3)}_n$ according to \eqref{eqn:d_Penalty_a_extension_map} analytically if $P\leq 2$ and numerically if $P\geq 3$.
\STATE \textbf{If} $d^{(3)}_n \neq 0$
\STATE \quad Update $\alpha_{n}=\alpha_{n}+d^{(3)}_n$.
\STATE \quad Update $\boldsymbol\Sigma^{(1)-1}_{\boldsymbol\alpha} = \boldsymbol \Sigma^{(1)-1}_{\boldsymbol\alpha}-d^{(3)}_ng_n
\boldsymbol \Sigma_{\boldsymbol\alpha}^{(1)-1}\mathbf S_{n}
(\mathbf{I}_P+ d^{(3)}_ng_n\mathbf S_{n}^H\boldsymbol\Sigma^{(1)-1}_{\boldsymbol\alpha}\mathbf S_{n})^{-1}\mathbf S_{n}^H\boldsymbol\Sigma^{(1)-1}_{\boldsymbol\alpha}$.
\STATE \textbf{end}
\ENDFOR
\STATE \textbf{until} $\boldsymbol\alpha$ satisfies some stopping criterion.
\end{algorithmic}\label{alg:MAP_descend_non_extension}
\end{algorithm}
\vspace{-3mm}
\subsection{MAP Estimation-based Device Activity Detection for Group Virtual Devices}
In this part, based on the signal model for virtual devices in \eqref{eqn:received_signal_time_xx}, we propose two MAP estimation-based device activity detection methods which detect  the
activities of the $N$ actual devices by detecting the activities of the $NP$ virtual devices.
First, we formulate an MAP estimation problem for activity detection of the $NP$ virtual devices. By \eqref{eq:equality_constraints}, \eqref{eqn_beta_n}, \eqref{eq:construct_alpha},  and \eqref{eqn:pmf_a_typical_a}, the p.m.f. of $\boldsymbol\beta$ is given by:
\begin{align}
  \begin{cases}
  \bar{p}(\boldsymbol\beta), & \boldsymbol\beta \text{ satisfies } \eqref{eq:equality_constraints},\\
  0, & \text{ otherwise},
  \end{cases}\label{eq:barbeta}
\end{align}
where
\vspace{-2mm}
\begin{align}
\bar{p}(\boldsymbol\beta)\triangleq\exp\bigg(\sum_{\omega\in\Psi}\bigg(c_{\omega}\prod_{n\in\omega}
\frac{\sum_{p\in\mathcal{P}}\beta_{(n-1)P+p}}{P}\bigg)-s\bigg). \label{density_beta}
\end{align}
Based on the conditional p.d.f. of $\mathbf{R}$ given $\boldsymbol\beta$ (identical to the likelihood function of $\mathbf{R}$ given in \eqref{eqn:likelihood_no_coop_xx}) and the prior distribution of $\boldsymbol\beta$, the conditional joint density of $(\mathbf{R},\boldsymbol\beta)$, for $\boldsymbol\beta$ satisfying \eqref{eq:equality_constraints}, is given by \cite{JiangSPAWC}:
 \begin{align}
 &p^{(4)}(\mathbf R,\boldsymbol\beta)
 \triangleq p^{(2)}(\mathbf R;\boldsymbol\beta)\bar{p}(\boldsymbol\beta)
= \frac{\exp\left(-{\rm tr}\left(\boldsymbol\Sigma^{(2)-1}_{\boldsymbol\beta}\mathbf R\mathbf R^H\right)+\sum\limits_{\omega\in\Psi}\left(c_{\omega}\prod\limits_{n\in\omega}\frac{\sum\limits_{p\in\mathcal{P}}\beta_{(n-1)P+p}}{P}\right)-s
 \right)}{\pi^{LM}\vert\boldsymbol\Sigma^{(2)-1}_{\boldsymbol\beta}\vert^M}. \label{eqn:likelihood_no_coop_map_extension}
 \end{align}
The maximization of $p^{(4)}(\mathbf R,\boldsymbol\beta)$ is equivalent to the minimization of $f^{(4)}(\boldsymbol\beta)$, where
\begin{align}\label{eqn:f_map}
  f^{(4)}(\boldsymbol\beta) \triangleq -\log p^{(4)}(\mathbf R,\boldsymbol\beta)-L\log\pi= f^{(2)}(\boldsymbol\beta)-\frac{1}{M}\sum_{\omega\in\Psi}\left(c_{\omega}\prod_{n\in\omega}
  \frac{\sum\limits_{p\in\mathcal{P}}\beta_{(n-1)P+p}}{P}\right).
\end{align}
Here, $f^{(2)}(\boldsymbol\beta)$ is given by \eqref{eqn:f_ml_ex}.
Thus, the MAP estimation problem of $\boldsymbol\beta$  from $\mathbf{R}$ given by \eqref{eqn:received_signal_time_xx}  can be formulated as follows.
\begin{Prob}[MAP Estimation for  Activity Detection of Virtual Devices]\label{prob:map_extension}
\begin{align*}
\min_{\boldsymbol\beta} \quad &f^{(4)}(\boldsymbol\beta) \\
s.t. \quad & \eqref{eq:equality_constraints}, \ \eqref{eqn_beta_n}.
\end{align*}
\end{Prob}

Similarly, the fact that $f^{(4)}(\boldsymbol\alpha)-f^{(2)}(\boldsymbol\alpha)$ decreases with $M$ indicates that the prior distribution of $\boldsymbol\beta$ becomes less important, and the observations $\mathbf{R}$ become more important, as $M$ increases. Consequently,  the MAP estimation given by Problem~\ref{prob:map_extension} is effective when $M$ is not large, and $\bar{p}(\boldsymbol\beta)$ can be obtained, compared to the ML estimation given by Problem~\ref{Prob:ML_a}.
Problem~\ref{prob:map_extension} is also a non-convex optimization problem over a convex set. It differentiates from Problem~\ref{prob:map}, as $\boldsymbol\Sigma^{(1)}_{\boldsymbol\alpha}$ and $\boldsymbol\Sigma^{(2)}_{\boldsymbol\beta}$ have different forms. Besides, the objective function of Problem~\ref{prob:map_extension} shares the same form as the objective function of the MAP estimation problem for device activity detection of the $NP$ devices under flat Rayleigh fading in \cite{JiangSPAWC} except that the dimensions of $\mathbf{S}\in\mathbb{C}^{L\times NP}$, $\mathbf{B}\in\mathbb{C}^{NP \times NP}$, $\mathbf{G}\in\mathbb{C}^{NP\times NP}$, and
$\mathbf{h}_m\in\mathbb{C}^{NP}$ rely on $P$. However, unlike Problem~\ref{prob:map} and the MAP estimation problem in \cite{JiangSPAWC}, Problem~\ref{prob:map_extension} has extra coupling constraints in \eqref{eq:equality_constraints} and hence cannot be directly solved by
the coordinate descent method. As in Section~\ref{sec:map_virtual}, to address this issue, we apply the penalty method to obtain a stationary point of an equivalent problem of Problem~\ref{prob:map_extension} in Section~\ref{sub_map_penal} and adopt relaxation to obtain a stationary point of an approximate problem of Problem~\ref{prob:map_extension} in Section~\ref{sub_map_relax}, respectively. Similarly, we shall see in Section~\ref{sec:map_virtual} that the device activity detection method via the penalty method has higher accuracy and higher computational complexity than the method via relaxation. Note that after solving Problem~\ref{prob:map_extension} for $\boldsymbol\beta$, we can construct the activities of the $N$ actual devices, $\boldsymbol\alpha$, according to \eqref{eq:construct_alpha}.
\subsubsection{Activity Detection for Virtual Devices via Penalty Method}\label{sub_map_penal}
First,
we disregard  the coupling
constraints in \eqref{eq:equality_constraints} and add a penalty for violating them to the objective function of Problem~\ref{prob:map_extension}.
Then, we can convert Problem~\ref{prob:map_extension} to the following problem.
\begin{Prob}[Penalty Problem of Problem~\ref{prob:map_extension}]\label{Prob:penal_extension}
\begin{align*}
\min_{\boldsymbol\beta}\quad & \tilde{f}^{(4)}(\boldsymbol\beta)\triangleq f^{(4)}(\boldsymbol\beta)
+\rho \eta(\boldsymbol\beta)\notag \\
s.t. \quad &  \eqref{eqn_beta_n},
\end{align*}
where $\rho>0$ is given by \eqref{eq:penaltyfunction}.
\end{Prob}

If $\rho$ is sufficiently large, an optimal solution of Problem~\ref{Prob:penal_extension} is also optimal for Problem~\ref{prob:map_extension} (as $f^{(4)}(\boldsymbol\beta)$ is bounded from above) \cite{Bertsekas99}.
Now, we can adopt the coordinate descent method to obtain a stationary point of Problem~\ref{Prob:penal_extension} instead of
Problem~\ref{prob:map_extension}. Specifically, given $\boldsymbol\beta$ obtained in the previous step, the coordinate optimization with respect to $\beta_i$ is equivalent to the optimization of the increment $d$ in $\beta_i$:
\begin{align}
\min_{d\in[-\beta_i,1-\beta_i]} \ \tilde{f}^{(4)}(\boldsymbol\beta + d\mathbf e_i).\label{eqn:Penal_MAP_a}
\end{align}
We shall see that it is more challenging to solve the problem in \eqref{eqn:Penal_MAP_a} for $P\in\{2,3,...\}$ than for $P = 1$.  Similarly, we define two basic functions before solving the problem:
\begin{align}
& \bar{\epsilon}_i(\boldsymbol\beta) \triangleq \sum_{\omega\in\Psi:n\in\omega}\bigg(c_{\omega}\prod_{\lceil \frac{i}{P} \rceil\in\omega,\lceil \frac{i}{P} \rceil\neq n}\frac{\sum\limits_{p\in\mathcal{P}}\beta_{(\lceil \frac{i}{P} \rceil-1)P+p}}{P}\bigg), \notag \\
& \tilde{f}^{(4)}_{\boldsymbol\beta,i}(d) \triangleq f^{(2)}_{\boldsymbol\beta,i}(d) - \frac{d}{M}\bar{\epsilon}_i(\boldsymbol\beta),\quad
\tilde{g}^{(4)}_{\boldsymbol\beta,i}(d) = A^{(4)}_id^3+B^{(4)}_id^2+C^{(4)}_id+D^{(4)}_i, \label{eq:three_times}
\end{align}
where
\begin{align*}
& A_i^{(4)}\triangleq A_i^{(2)}, \quad B_i^{(4)}\triangleq B_i^{(2)} - \frac{\delta_i^2\bar{\epsilon}_i(\boldsymbol\beta)}{M} \bigg(\mathbf S_{:,i}^H\mathbf \Sigma^{(2)-1}_{\boldsymbol\beta}\mathbf S_{:,i}\bigg)^2, \\
&C_i^{(4)}\triangleq C_i^{(2)} - 2\delta_i \bar{\epsilon}_i(\boldsymbol\beta) \bigg(\mathbf S_{:,i}^H\mathbf \Sigma^{(2)-1}_{\boldsymbol\beta}\mathbf S_{:,i}\bigg),\quad
 D_i^{(4)}\triangleq D_i^{(2)}-\bar{\epsilon}_n(\boldsymbol\beta).
\end{align*}
Note that $\tilde{g}^{(4)}_{\boldsymbol\beta,i}(d)$ is the numerator of the derivative of $\tilde{f}^{(4)}_{\boldsymbol\beta,i}(d)$ (which is a fraction, as shown in Appendix D). The optimal solution of the problem in~\eqref{eqn:Penal_MAP_a} is summarized below.
\begin{Thm}[Optimal Solutions of Coordinate Optimization in \eqref{eqn:Penal_MAP_a}]\label{Thm:Step_APs_Penal_M1}
Given $\boldsymbol\beta$ obtained in the previous step, the optimal solution of the problem in~\eqref{eqn:Penal_MAP_a} is given by:
\begin{align}
d^{(4)}_i = \arg\min\limits_{d\in\mathcal{D}^{(4)}_i\cup\{-\beta_i,1-\beta_i\}} \tilde{f}^{(4)}_{\boldsymbol\beta,i}(d),\label{eqn:d_Penalty_a_MAP}
\end{align}
where $\mathcal{D}^{(4)}_i \triangleq \{d\in[-\beta_i,1-\beta_i]:g^{(4)}_{\boldsymbol\beta,i}(d) = 0\}$.
\end{Thm}
\begin{IEEEproof}
Please refer to Appendix D.
\end{IEEEproof}

As $\tilde{g}^{(4)}_{\boldsymbol\beta,i}(d)$ is a polynomial with degree 3, $\mathcal{D}^{(4)}_i$ can be obtained in closed-form with computational complexity $\mathcal{O}(P+2^N)$.
The details of the coordinate descent method are summarized in Algorithm~\ref{alg:MAP_descend_2}.
\begin{algorithm}[t] \caption{\small{Algorithm for Solving Problem~\ref{Prob:penal_extension} and Problem~\ref{Prob ML_relax_xx}}}
\scriptsize
\hspace*{0.02in} {\bf Input:}
empirical covariance matrix $\widehat{\mathbf \Sigma}_{\mathbf R}$. \\
\hspace*{0.02in} {\bf Output:}
activities of virtual devices $\boldsymbol\beta$.
\begin{algorithmic}[1]
\STATE Initialize $\mathbf \Sigma^{(2)-1}_{\boldsymbol\beta}=\frac{1}{\sigma^2} \mathbf I_L$, $\boldsymbol\beta=\mathbf 0$.
\STATE \textbf{repeat}
\FOR {$i\in\mathcal{I}$}
\STATE {\bf Penalty Method for Solving Problem~\ref{Prob:penal_extension}:} Calculate $d^{(4)}_i$ according to \eqref{eqn:d_Penalty_a_MAP}. \\
{\bf Relaxation Method for Solving Problem~\ref{Prob ML_relax_xx}:} Calculate $d^{(4)}_i$ according to \eqref{eqn:d_MAP_a_2}.
\STATE \textbf{If} $d^{(4)}_i\neq 0$
\STATE \quad Update $\beta_{i}=\beta_{i}+d^{(4)}_i$.
\STATE \quad Update $\mathbf \Sigma^{(2)-1}_{\boldsymbol\beta} = \mathbf \Sigma^{(2)-1}_{\boldsymbol\beta}-\frac{d^{(4)}_i\delta_i\mathbf \Sigma_{\boldsymbol\beta}^{(2)-1}\mathbf S_{:,i}\mathbf S_{:,i}^H\mathbf \Sigma^{(2)-1}_{\boldsymbol\beta}}{1+d^{(4)}_i\delta_i\mathbf S_{:,i}^H\mathbf \Sigma^{(2)-1}_{\boldsymbol\beta}\mathbf S_{:,i}}$.
\STATE \textbf{end}
\ENDFOR
\STATE \textbf{until} $\boldsymbol\beta$ satisfies some stopping criterion.
\end{algorithmic}\label{alg:MAP_descend_2}
\end{algorithm}
If each problem in \eqref{eqn:Penal_MAP_a} has a unique optimal solution, Algorithm~\ref{alg:MAP_descend_2} for solving Problem~\ref{Prob:penal_extension}  converges to a stationary point of Problem~\ref{Prob:penal_extension}, as the number of iterations goes to infinity \cite[Proposition 2.7.1]{Bertsekas99}.
The computational complexities of Step 4, Step 6, and Step 7 are $\mathcal{O}(L^2+2^N)$, $\mathcal{O}(1)$, and $\mathcal{O}(L^2+2^N)$, respectively, as $N,L\rightarrow\infty$. Thus, the computational complexity of each iteration of Algorithm \ref{alg:MAP_descend_2} for solving Problem~\ref{Prob:penal_extension}  is $\mathcal{O}\left(NPL^2+NP2^N\right)$,  higher than that for solving Problem~\ref{Prob:penal}. As illustrated in Section~\ref{sec:map_actual}, the actual computational complexity can be much lower.
\subsubsection{Activity Detection for Virtual Devices via Relaxation}\label{sub_map_relax}
First, we relax the coupling constraints in \eqref{eq:equality_constraints} of Problem~\ref{prob:map_extension} as in Section~\ref{subsec:relax}. The relaxed version of Problem~\ref{prob:map_extension} is given as follows.
\begin{Prob}[Relaxed Problem of Problem~\ref{prob:map_extension}]\label{Prob ML_relax_xx}
\begin{align}
\min_{\boldsymbol\beta} &\quad f^{(4)}(\boldsymbol\beta)\notag\\
s.t.  & \quad \eqref{eqn_beta_n}.\notag
\end{align}
\end{Prob}
\vspace{-2mm}
Obviously, Problem~\ref{Prob ML_relax_xx} has the same variables and constraints but a simpler objective function than Problem~\ref{Prob:penal_extension}.
Thus, we expect that Problem~\ref{Prob ML_relax_xx} can be solved with lower computational complexity than Problem~\ref{Prob:penal_extension}. Problem~\ref{Prob ML_relax_xx} shares the same form as the MAP estimation problem for activity
detection of the $NP$ devices under flat Rayleigh fading in \cite{JiangSPAWC}.
Thus, the coordinate descent method in \cite[Algorithm 1]{JiangSPAWC} can be directly used to obtain a stationary point of Problem~\ref{Prob ML_relax_xx}. We present the coordinate descent method here for completeness. Specifically, given $\boldsymbol\beta$ obtained in the previous step, the coordinate optimization with respect to $\beta_i$ is equivalent to the optimization of the increment $d$ in $\beta_i$:
\begin{align}
\min_{d\in [-\beta_i,1-\beta_i]} \ f^{(4)}(\boldsymbol\beta + d\mathbf e_i).\label{eqn:Penal_a_relax_xx}
\end{align}
Define
\begin{align*}
f_{\boldsymbol\beta,i}^{(4)}(d) \triangleq
 \log(1+d\delta_i\mathbf S_{:,i}^H\mathbf \Sigma^{(2)-1}_{\boldsymbol\beta}\mathbf S_{:,i}) -\frac{d\delta_i\mathbf S_{:,i}^H\mathbf \Sigma^{(2)-1}_{\boldsymbol\beta}\widehat{\mathbf \Sigma}_{\mathbf R}\mathbf \Sigma^{(2)-1}_{\boldsymbol\beta}\mathbf S_{:,i} }{1+d\delta_i\mathbf S_{:,i}^H\mathbf \Sigma^{(2)-1}_{\boldsymbol\beta}\mathbf S_{:,i}} - \frac{d\bar{\epsilon}_i(\boldsymbol\beta)}{M}.
\end{align*}
Then, the optimal solution of the problem in \eqref{eqn:Penal_a_relax_xx} is given by \cite{JiangSPAWC}:
\begin{align}
d^{(4)}_i =
\begin{cases}
\min\left\{\max\left\{\textcolor{black}{s_i(\boldsymbol\beta)},-\beta_i\right\}, 1-\beta_i\right\},& C_i\leq 0\\
\arg\min_{d\in\{s_i(\boldsymbol\beta),-\beta_i+1\}} \bar{f}_{\boldsymbol\beta,i}^{(4)}(d), & C_i>0, \Delta_i>0\\
-\beta_i+1,& C_i>0, \Delta_i\leq 0
\end{cases}
,\label{eqn:d_MAP_a_2}
\end{align}
where $\textcolor{black}{s_i(\boldsymbol\beta)} \triangleq \frac{1}{2C_i}(1-\sqrt{\Delta_i})-\frac{1}{\delta_i\mathbf S_{:,i}^H\mathbf \Sigma_{\boldsymbol\beta}^{(2)-1}\mathbf S_{:,i}}$, $C_i\triangleq \frac{\bar{\epsilon}_i(\boldsymbol\beta)}{M}$, and $\Delta_i \triangleq 1-\frac{4C_i\mathbf S_{:,i}^H\mathbf \Sigma^{(2)-1}_{\boldsymbol\beta}\widehat{\mathbf \Sigma}_{\mathbf R}\mathbf \Sigma^{(2)-1}_{\boldsymbol\beta}\mathbf S_{:,i}}{\delta_i(\mathbf S_{:,i}^H\mathbf \Sigma_{\boldsymbol\beta}^{(2)-1}\mathbf S_{:,i})^2}$. Note that in the i.i.d. case where $\alpha_n,n\in\mathcal{N}$ are i.i.d. with $\text{Pr}\left[\alpha_n = 1\right] = q$, $\bar{p}(\boldsymbol\beta)$ in \eqref{density_beta} becomes:
\begin{align}
  \bar{p}(\boldsymbol\beta)= \text{exp}\left(\log\frac{q}{1-q}\sum_{n\in\mathcal{N}}
 \frac{\sum_{p\in\mathcal{P}}\beta_{(n-1)P+p}}{P}+N\log(1-q)\right). \label{eq:mvb_iidmap}
\end{align}
The optimal solution of the problem in \eqref{eqn:Penal_a_relax_xx} for the i.i.d. case is given by \cite{JiangSPAWC}:
{\scriptsize{
\begin{align}
 & d^{(4)}_i = \min\left\{\max\left\{
  \frac{MP}{2\log\frac{q}{1-q}}
  \left(1-\sqrt{1-\frac{4\log\frac{q}{1-q}
  \mathbf{S}_{:,i}^H\boldsymbol\Sigma^{(2)-1}_{\boldsymbol\beta}
  \widehat{\mathbf \Sigma}_{\mathbf R}
  \boldsymbol\Sigma^{(2)-1}_{\boldsymbol\beta}
  \mathbf{S}_{:,i}}{MP\delta_i(\mathbf{S}_{:,i}^H\boldsymbol\Sigma^{(2)-1}_{\boldsymbol\beta}
  \mathbf{S}_{:,i})^2}}\right)-\frac{1}{\delta_i\mathbf{S}_{:,i}^H\boldsymbol\Sigma^{(2)-1}_{\boldsymbol\beta}
  \mathbf{S}_{:,i}},-\beta_i\right\},1-\beta_i\right\}. \label{eq:relax_map}
\end{align}}}
\vspace{-2mm}

From \eqref{eq:relax_map}, we can see that as $M\to \infty$ or $p_{\alpha}\to 0.5$,
the optimal solution in \eqref{eq:relax_map} reduces to the optimal solution  in \eqref{eqn:d_ML_relax}. In addition, as $p_{\alpha}\to 0$,
the optimal solution in \eqref{eq:relax_map}  becomes $-\beta_i$, and hence the updated $\beta_i$ converges to $0$.
The details of the coordinate descent method for solving Problem~\ref{Prob ML_relax_xx} are also summarized in Algorithm~\ref{alg:MAP_descend_2}.  If each problem in \eqref{eqn:Penal_a_relax_xx} has a unique optimal solution, Algorithm~\ref{alg:MAP_descend_2} for solving  Problem~\ref{Prob ML_relax_xx} converges to a stationary point of Problem~\ref{Prob ML_relax_xx}, as the number of iterations goes to infinity \cite[Proposition 2.7.1]{Bertsekas99}. The computational complexities of Step 4, Step 6, and Step 7 are $\mathcal{O}(L^2+2^N)$, $\mathcal{O}(1)$, and $\mathcal{O}(L^2+2^N)$, respectively, as $N,L\rightarrow\infty$. Thus, the computational complexity of each iteration of Algorithm \ref{alg:MAP_descend_2} for solving  Problem~\ref{Prob ML_relax_xx} is $\mathcal{O}\left(NPL^2+NP2^N\right)$. Similarly, the actual computational complexity can be much lower.
\vspace{-4mm}
\section{Numerical Results}\label{sec:simulation}
In this section, we evaluate the performance of the three proposed ML estimation-based device activity detection methods given by Algorithm~\ref{alg:ML_descend_non_extension} and Algorithm~\ref{alg:ML_descend}, referred to as {\em Prop.-ML-act.}, {\em Prop.-ML-virt.}, and {\em Prop.-ML-virt.-rel.}, respectively, and the three proposed MAP estimation-based device activity detection methods given by Algorithm~\ref{alg:MAP_descend_non_extension} and Algorithm~\ref{alg:MAP_descend_2}, referred to as {\em Prop.-MAP-act.}, {\em Prop.-MAP-virt.}, and {\em Prop.-MAP-virt.-rel.}, respectively.  It is noteworthy that device activity detection for massive grant-free access under frequency-selective Rayleigh fading
has not been studied so far. Thus, we construct five baseline schemes, namely, {\em BL-ML-act.}, {\em BL-GL-act.}, {\em BL-AMP-act.}, {\em BL-GL-virt.-rel.}, and {\em BL-AMP-virt.-rel.}, based on the existing device activity detection methods, i.e., ML estimation \cite{Yu19ICC}, GROUP LASSO \cite{JSAC_li}, and AMP \cite{Liu18TSP}, which are designed for flat Rayleigh fading, and the proposed signal model for virtual devices under frequency-selective Rayleigh fading in \eqref{eqn:received_signal_time_xx}. Specifically,  {\em BL-ML-act.}, {\em BL-GL-act.}, and {\em BL-AMP-act.} directly estimate the activities of the $N$ actual devices, $\boldsymbol\alpha$, from $\mathbf{r}_m,m\in\mathcal{M}$ given by \eqref{eqn:received_signal_time} via ignoring the difference between the statistics of $\mathbf{r}_m,m\in\mathcal{M}$ under flat Rayleigh fading and under frequency-selective Rayleigh fading and using ML estimation \cite{Yu19ICC}, GROUP LASSO \cite{JSAC_li}, and AMP \cite{Liu18TSP}, respectively.  On the other hand, {\em BL-GL-virt.-rel.} and {\em BL-AMP-virt.-rel.} first detect the activities of the $NP$ virtual devices, $\boldsymbol\beta$, from $\mathbf{r}_m,m\in\mathcal{M}$ given by \eqref{eqn:received_signal_time_xx},  using GROUP LASSO \cite{JSAC_li} and AMP \cite{Liu18TSP}, respectively, without considering the constraints in \eqref{eq:equality_constraints}, and then set the activities of the $N$ actual devices, $\boldsymbol\alpha$, according to \eqref{eq:construct_alpha}. The thresholds for the proposed methods, {\em BL-GL-act.}, and {\em BL-GL-virt.-rel.} are numerically optimized \cite{JSAC_li,Yu19ICC,Jiang21TWC}. The thresholds for {\em BL-AMP-act.} and {\em BL-AMP-virt.-rel.} are chosen as in \cite{Liu18TSP}.
Notice that the computational complexities of {\em BL-ML-act.}, {\em BL-GL-act.}, {\em BL-AMP-act.}, {\em BL-GL-virt.-rel.}, and {\em BL-AMP-virt.-rel.} are $\mathcal{O}(NPL^2)$, $\mathcal{O}(NL^2)$, $\mathcal{O}(NLM)$, $\mathcal{O}(NPL^2)$, and $\mathcal{O}(NPLM)$, respectively, as compared with those of the proposed methods given in Table~\ref{tab:computation_complexity}.

We consider two cases of device activity distributions, i.e., the i.i.d. case and correlated case. For a fair comparison, in both cases, the probability of each device being active is $q$. Specifically, in the i.i.d. case, the activities of devices are i.i.d. with probability $q$ of being active. As a result, $p(\boldsymbol\alpha)$ is given by $p(\boldsymbol\alpha) = \text{exp}\left(\log\frac{q}{1-q}\sum_{n\in\mathcal{N}}\alpha_n+N\log(1-  q)\right)$,
and $\bar{p}(\boldsymbol\beta)$ is given by \eqref{eq:mvb_iidmap}. For the correlated case, we consider the group activity model \cite{Jiang21TWC} where the devices in $\mathcal{N}$  are  divided into $K$ groups, the activities of devices in the same group are identical, and the group activities are i.i.d. with probability $q$ of being active. For all $k\in\mathcal{K}\triangleq\{1,2,...,K\}$, let $\mathcal N_k\subseteq\mathcal{N}$ denote the set of indices of the devices in the $k$-th group. Note that $\cup_{k\in\mathcal K}\mathcal N_k=\mathcal{N}$ and $\mathcal N_k\cap\mathcal N_{k'}=\emptyset$, for all $k,k'\in\mathcal K$, $k\neq k^{'}$.  Under the group activity model \cite{Jiang21TWC}, 
$p(\boldsymbol\alpha)$ is well approximated by \eqref{eqn:pmf_a_typical_a} with a small $\epsilon>0$, where $c_{\omega}=0, \omega\nsubseteq \mathcal N_k$ for all $k\in\mathcal{K}$, and 
\begin{align}
c_{\omega} = &
\begin{cases}
(-1)^{|\omega|}\log\left(\frac{1-q}{N\epsilon}\right), \quad & |\omega|<|\mathcal N_k|,\\
\log\left(\frac{q}{1-q}\right), & |\omega|=|\mathcal N_k|, |\omega|\text{ is odd},\\
\log\left(\frac{q(1-q)}{\epsilon^2}\right), & |\omega|=|\mathcal N_k|, |\omega|\text{ is even},\\
\end{cases}\quad  \omega\subseteq \mathcal N_k\text{ for some }k\in\mathcal K. \notag
\end{align}
In the simulation, we choose $\epsilon=10^{-3}$. Furthermore, in the simulation, we adopt the following setup. We generate pilots according to i.i.d. $\mathcal{CN}(0,\mathbf{I}_L)$ and normalize their norms to $\sqrt{L}$ \cite{JSAC_li,Caire18ISIT,Yu19ICC}.
We independently generate $1000$ realizations for $\alpha_n,n\in\mathcal N$, $ h_{n,m,p} \sim\mathcal{CN}(0,1)$, $n\in\mathcal N$, $m\in\mathcal M$, $p\in\mathcal P$, and Gaussian pilots and evaluate the average error rate over all $1000$ realizations. We set $N=1000$, $g_n=1,n\in\mathcal{N}$, and $\sigma^2=0.1$.
Unless otherwise stated, we choose $L=72$, $M=128$, $P=4$,  $K=2$, and $q=0.07$.

\begin{figure}[t]
\begin{center}
\subfigure[\scriptsize{i.i.d. case.
}\label{fig:error_vs_P_iid_BL}]
{\resizebox{5cm}{!}{\includegraphics{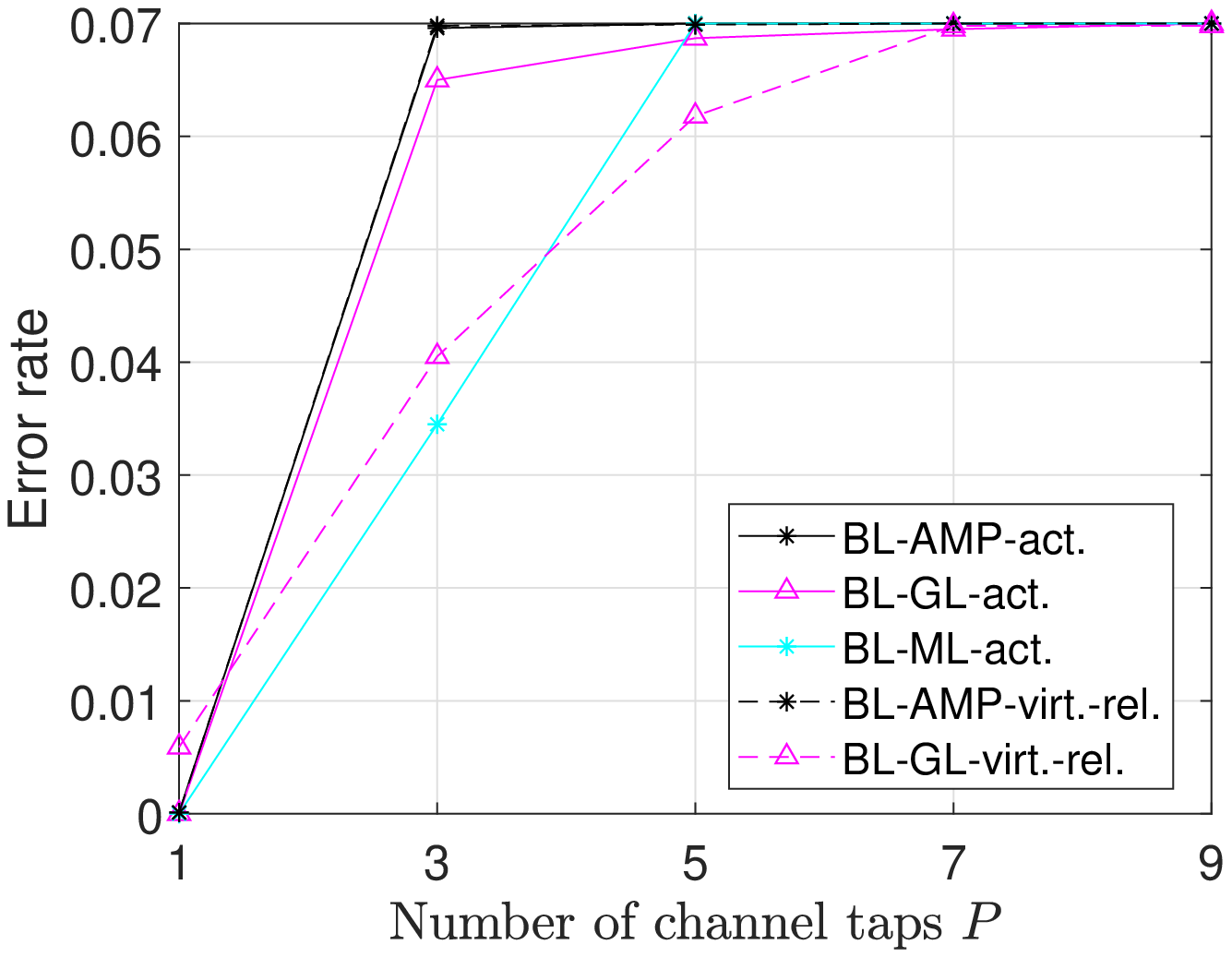}}}\quad
\subfigure[\scriptsize{Correlated case.
}\label{fig:error_vs_P_corr_BL}]
{\resizebox{5cm}{!}{\includegraphics{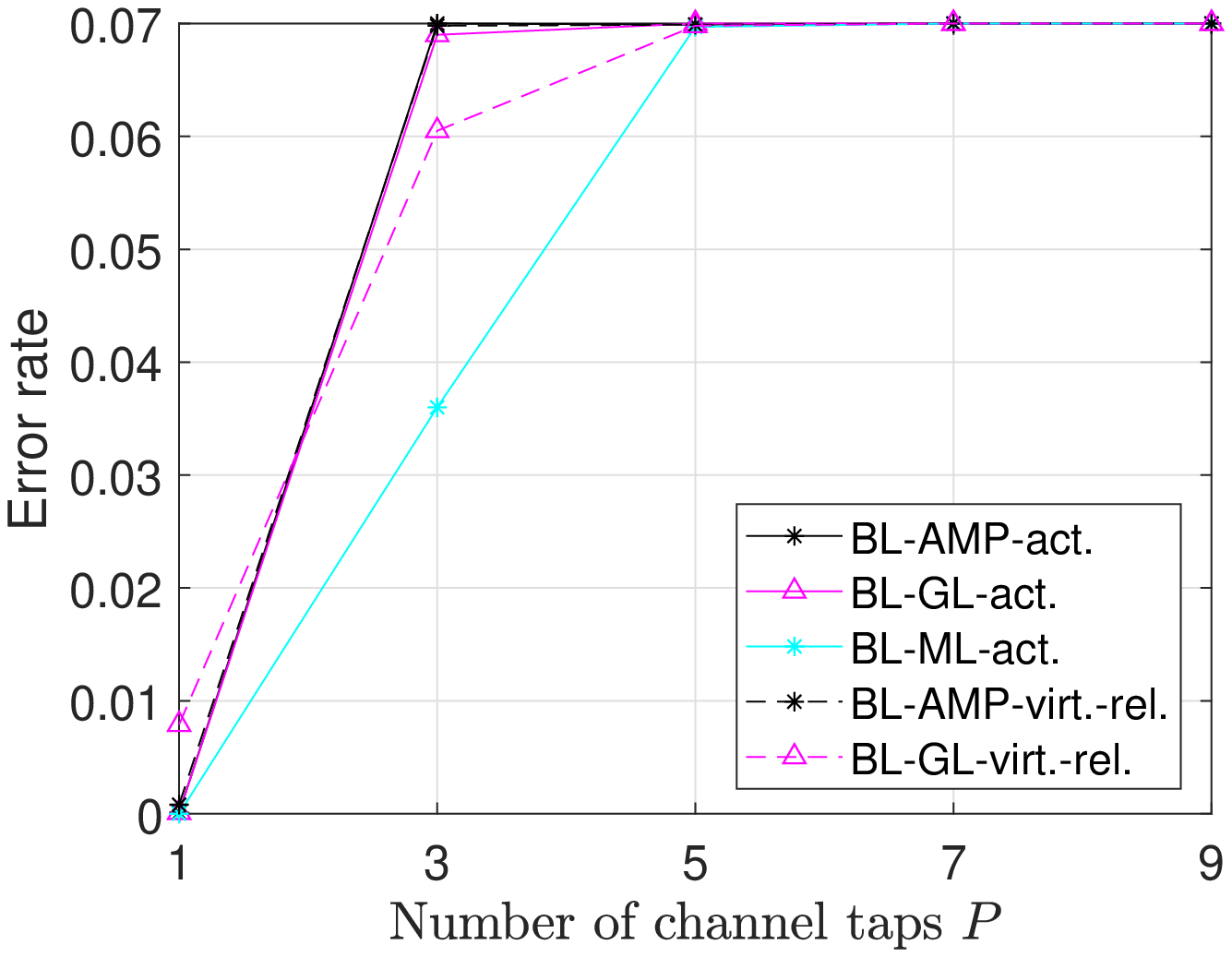}}}
\end{center}
\vspace{-4mm}
\caption{\small{Error rates of baseline schemes versus number of channel taps $P$.}}
\vspace{-6mm}
\label{fig:P_BL}
\end{figure}
\begin{figure}[t]
\begin{center}
\subfigure[\scriptsize{i.i.d. case.
}\label{fig:error_vs_P_iid}]
{\resizebox{5cm}{!}{\includegraphics{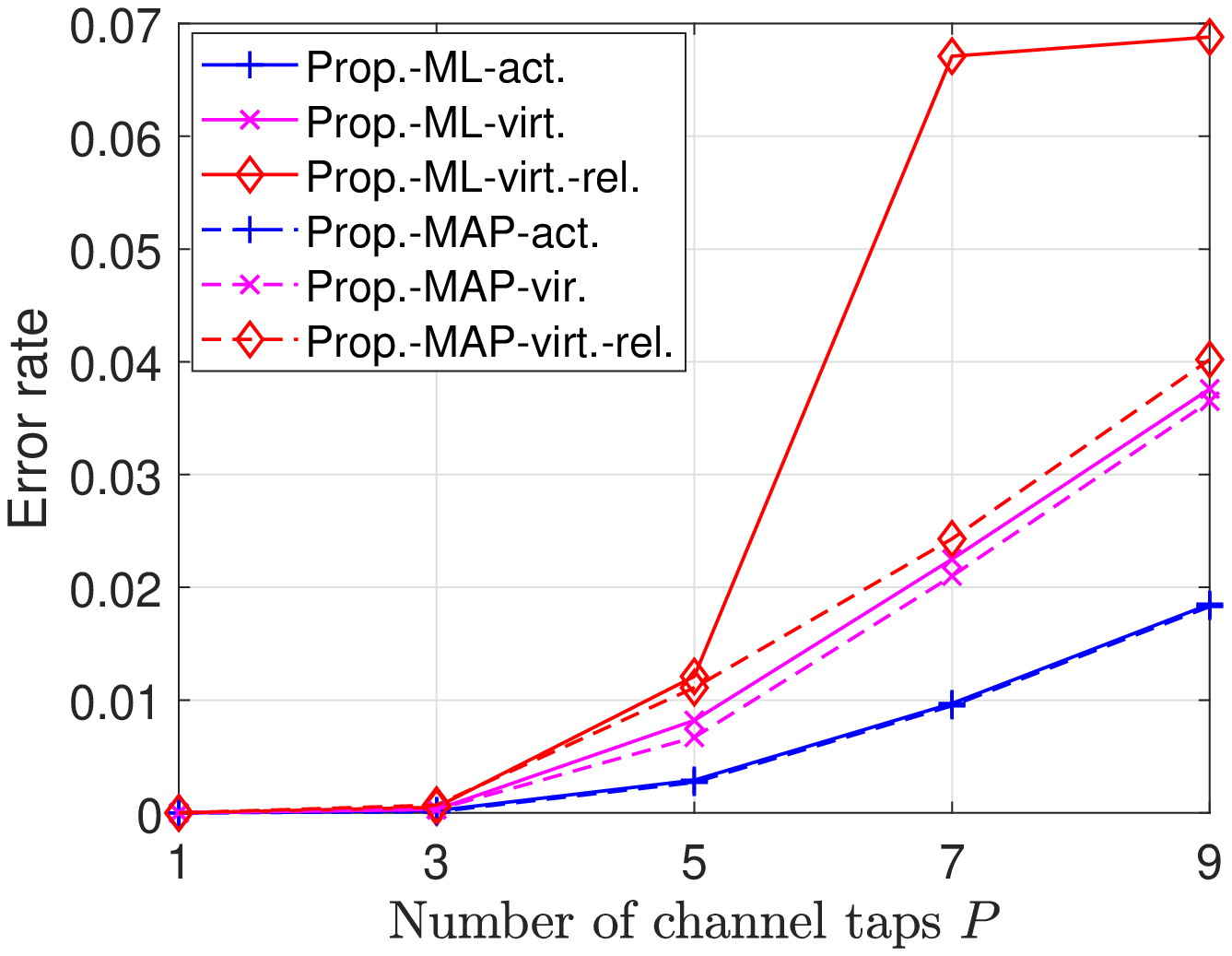}}}\quad
\subfigure[\scriptsize{Correlated case.
}\label{fig:error_vs_P_corr}]
{\resizebox{5cm}{!}{\includegraphics{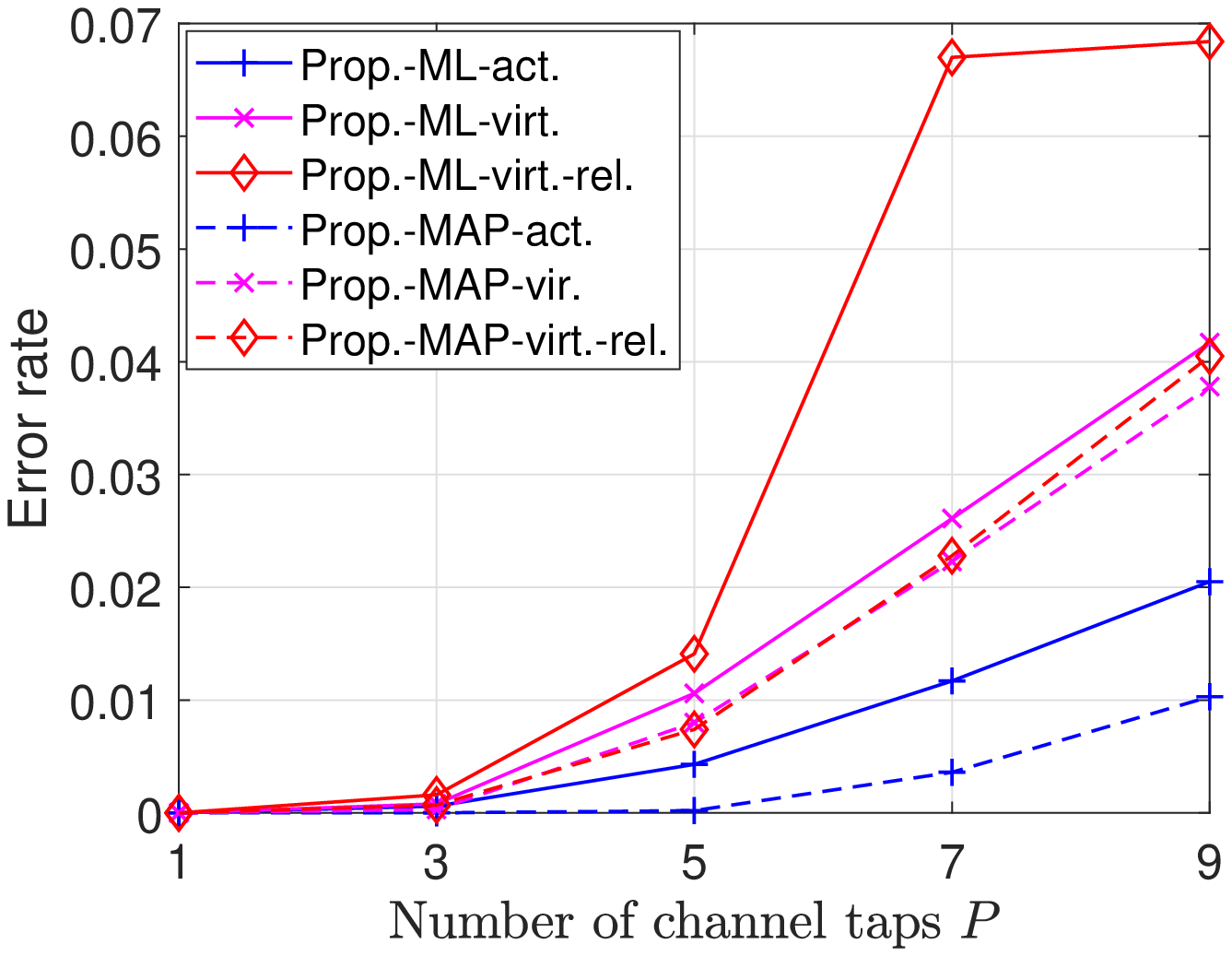}}}
\end{center}
\vspace{-4mm}
\caption{\small{Error rates of proposed solutions versus number of channel taps $P$.}}
\vspace{-7mm}
\label{fig:P}
\end{figure}
\begin{figure}[t]
\begin{center}
\subfigure[\scriptsize{i.i.d. case.
}\label{fig:error_vs_L_iid}]
{\resizebox{5cm}{!}{\includegraphics{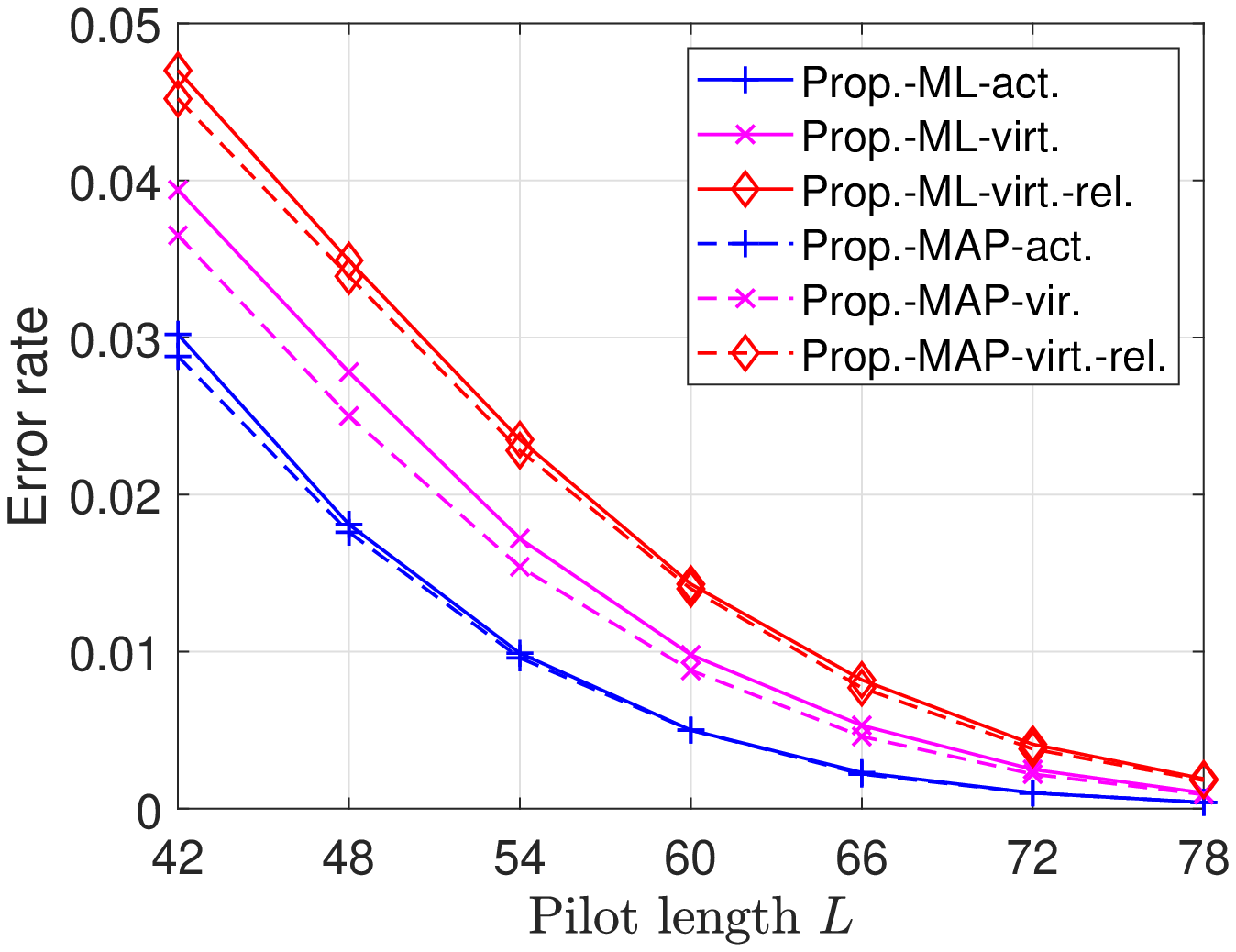}}}\quad
\subfigure[\scriptsize{Correlated case.
}\label{fig:error_vs_L_corr}]
{\resizebox{5cm}{!}{\includegraphics{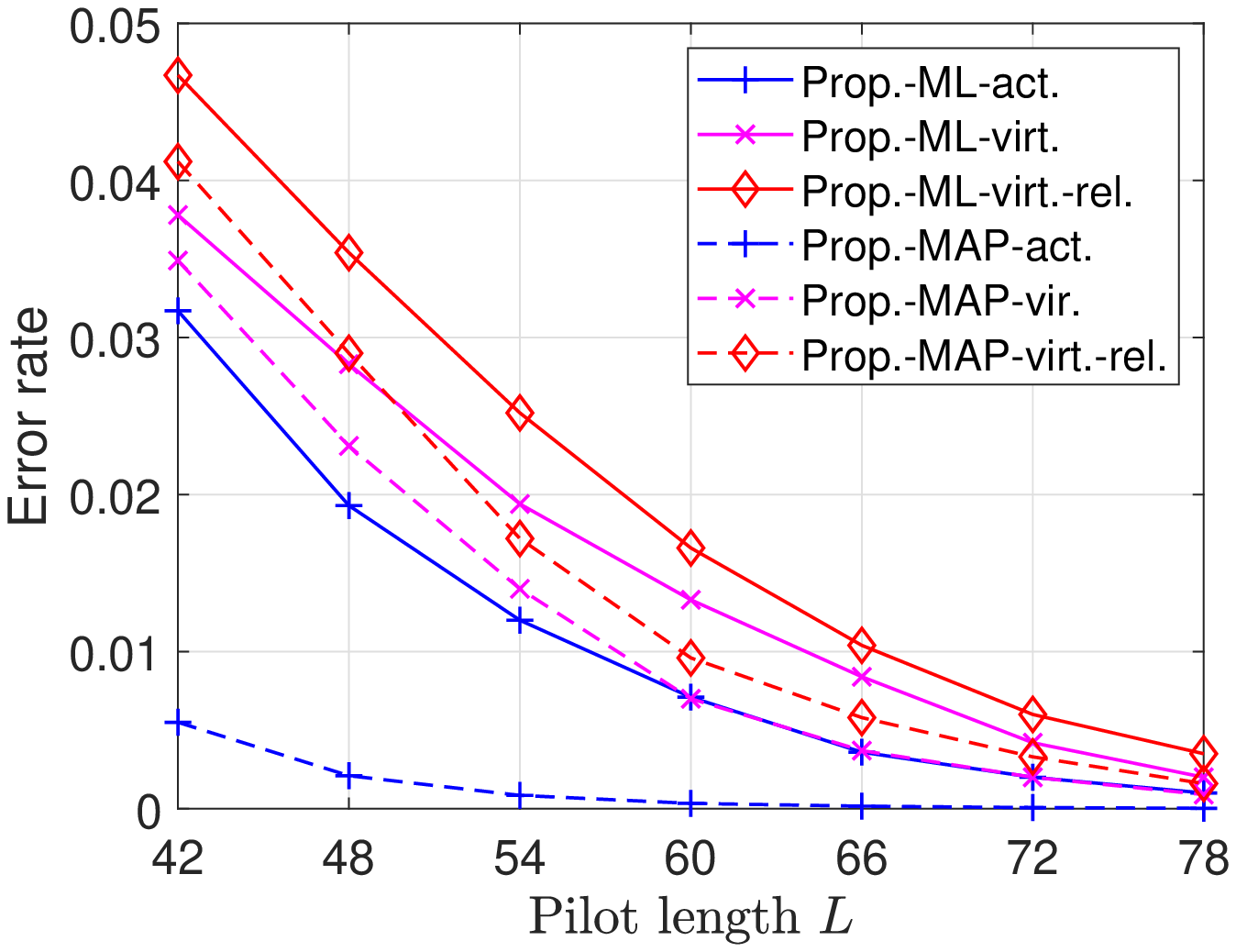}}}
\end{center}
\vspace{-4mm}
\caption{\small{Error rates of proposed solutions versus pilot length $L$.}}
\vspace{-7mm}
\label{fig:L}
\end{figure}
\begin{figure}[t]
\begin{center}
\subfigure[\scriptsize{i.i.d. case.
}\label{fig:error_vs_M_iid}]
{\resizebox{5cm}{!}{\includegraphics{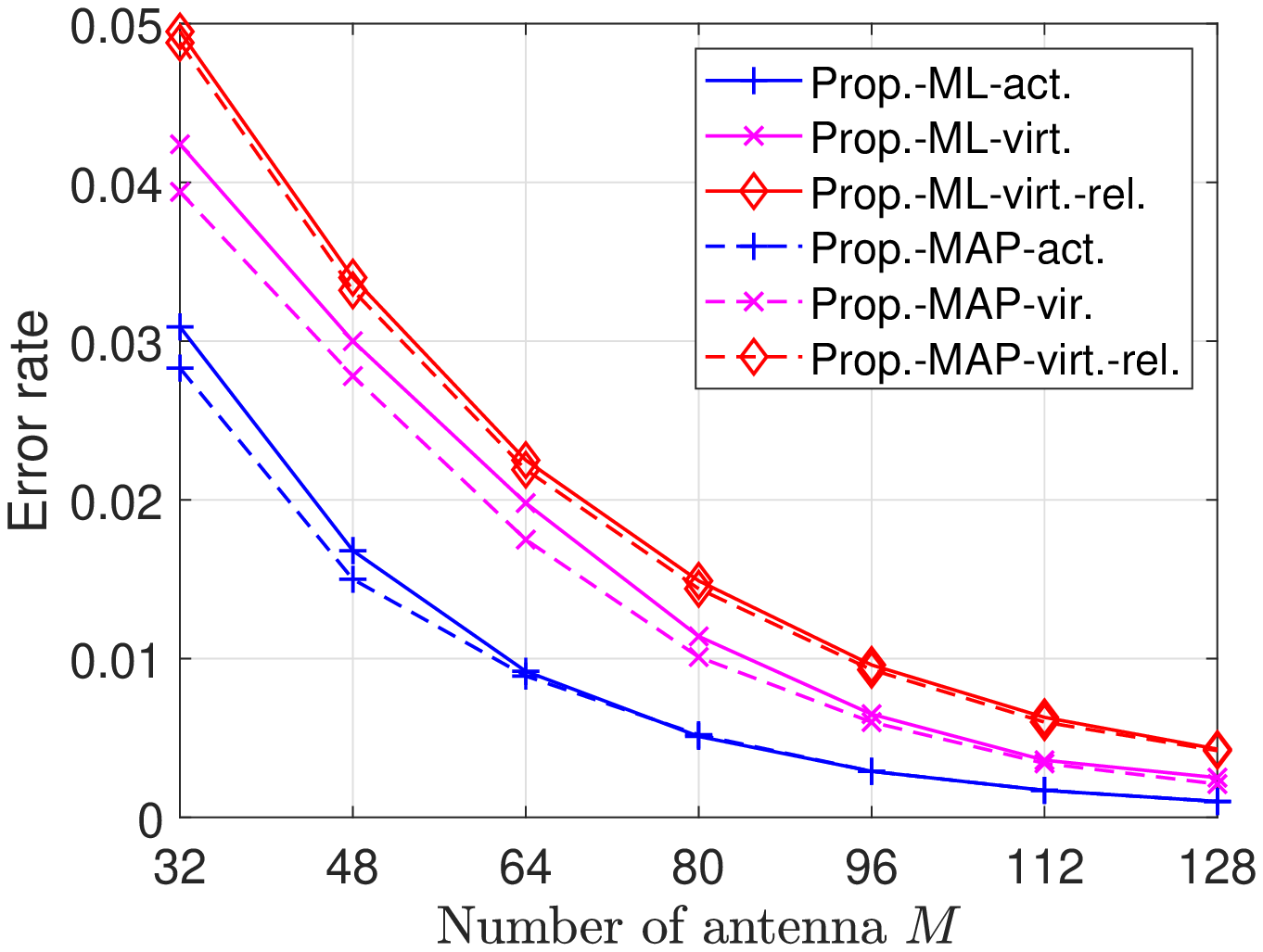}}}\quad
\subfigure[\scriptsize{Correlated case.
}\label{fig:error_vs_M_corr}]
{\resizebox{5cm}{!}{\includegraphics{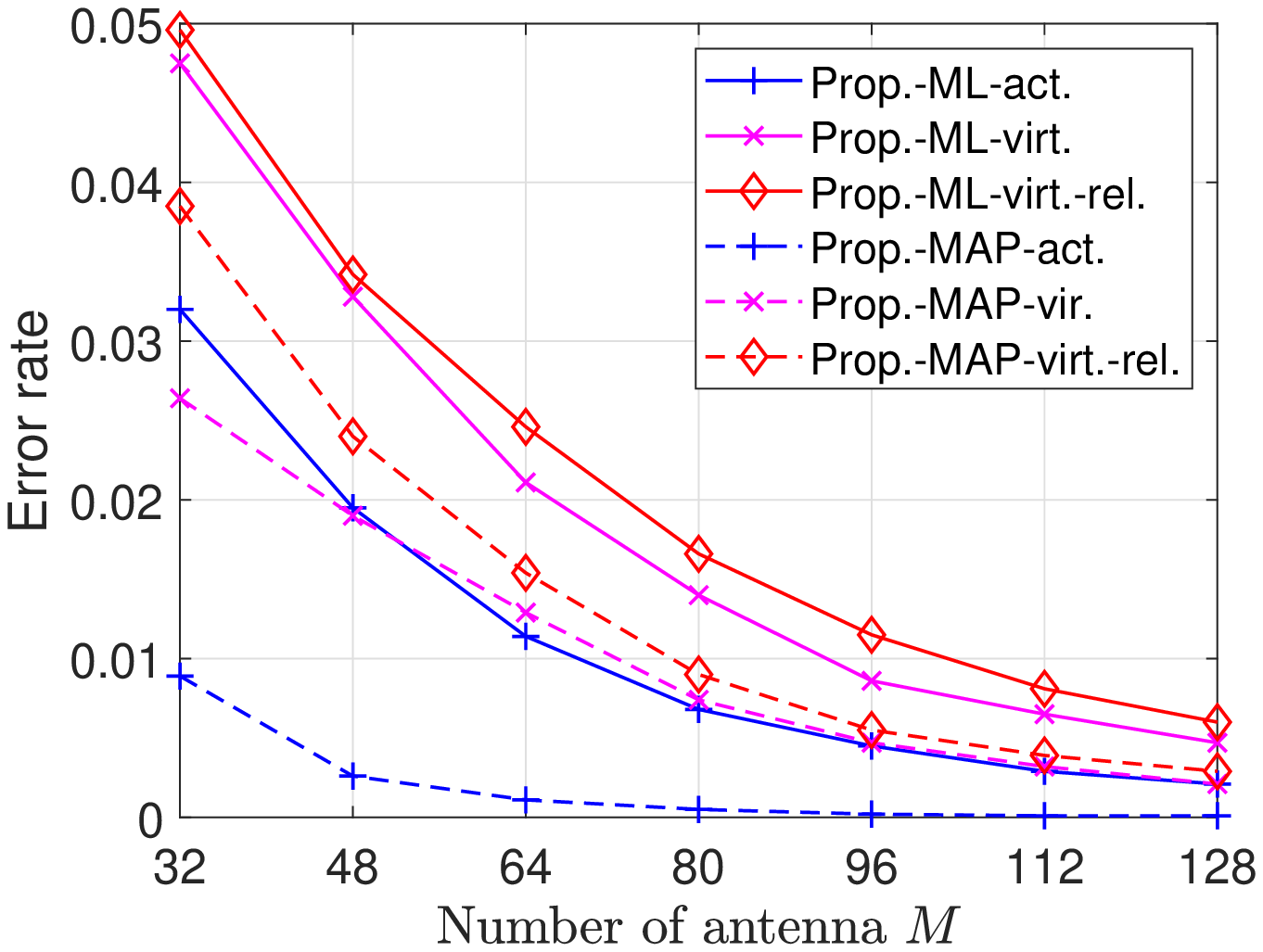}}}
\end{center}
\vspace{-4mm}
\caption{\small{Error rates of proposed solutions versus number of antennas $M$.}}
\vspace{-7mm}
\label{fig:M}
\end{figure}
\begin{figure}[t]
\begin{center}
\subfigure[\scriptsize{i.i.d. case.
}\label{fig:error_vs_K_iid}]
{\resizebox{5cm}{!}{\includegraphics{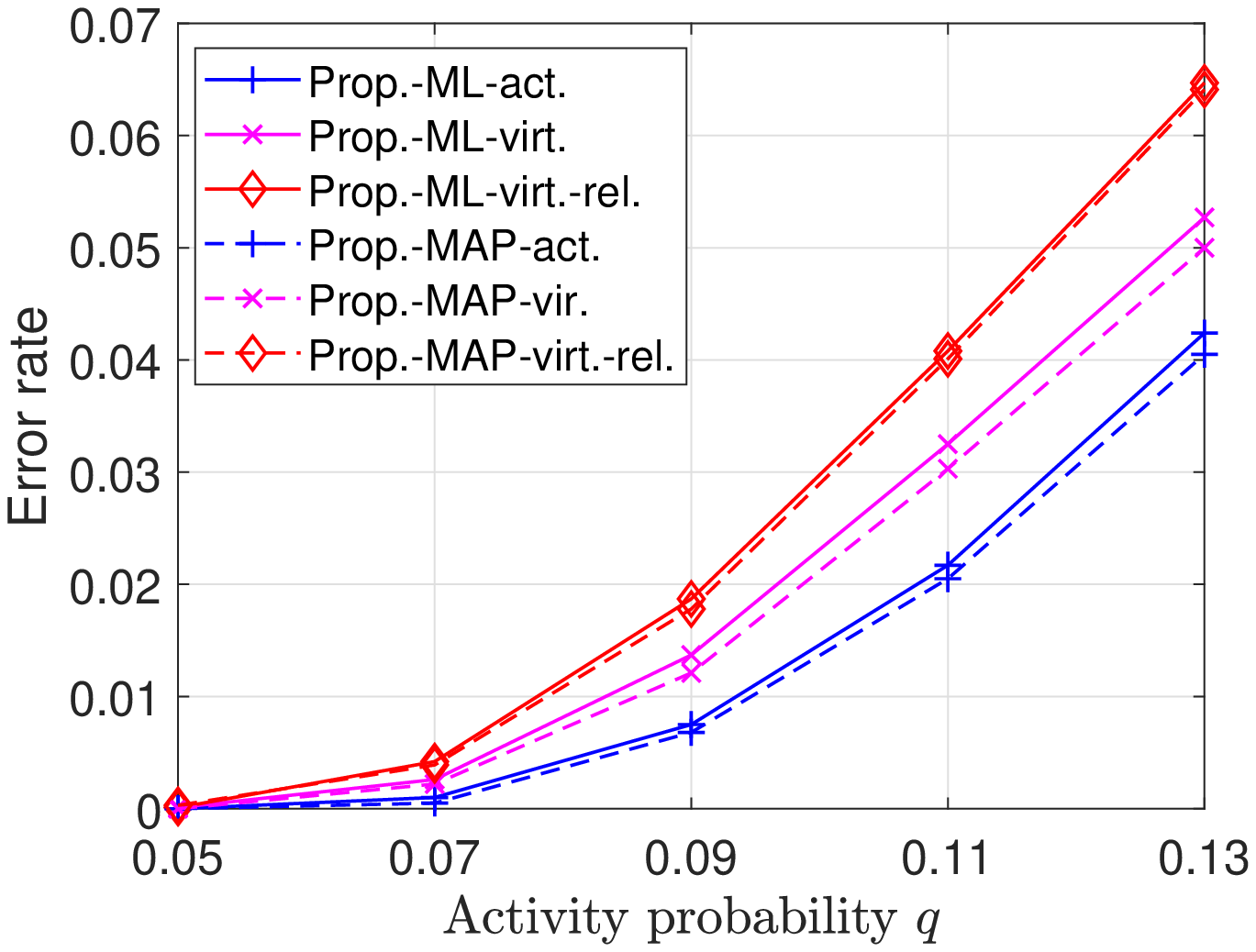}}}\quad
\subfigure[\scriptsize{Correlated case.
}\label{fig:error_vs_K_corr}]
{\resizebox{5cm}{!}{\includegraphics{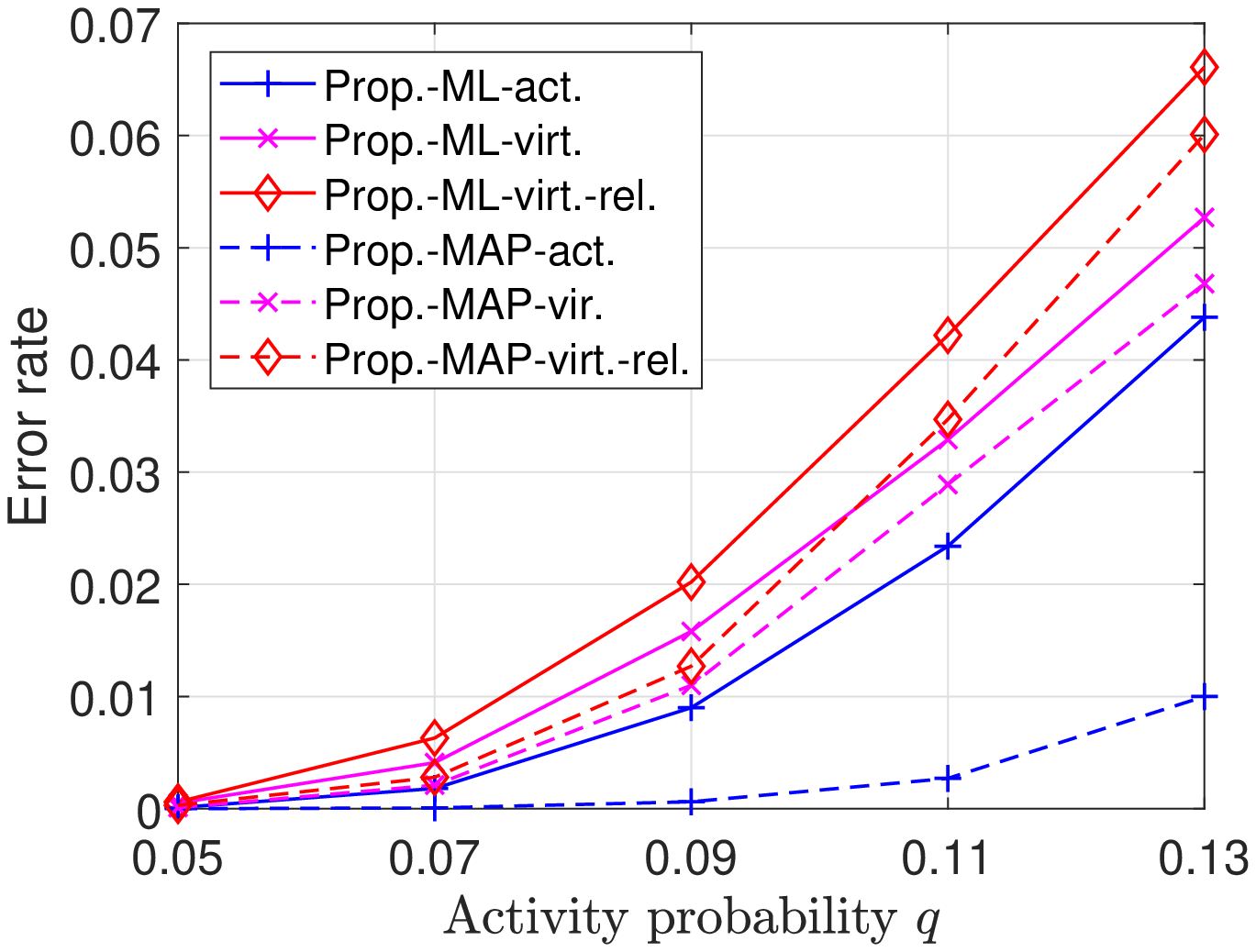}}}
\end{center}
\vspace{-4mm}
\caption{\small{Error rates of proposed solutions versus activity probability $q$.}}
\vspace{-9mm}
\label{fig:K}
\end{figure}
\begin{figure}[t]
\begin{center}
\includegraphics[width=5cm]{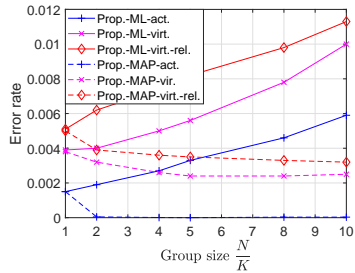}
\end{center}
\vspace{-5mm}
\caption{\small{Error rates of proposed solutions versus group size $\frac{N}{K}$ in correlated case.}}
\vspace{-7mm}
\label{fig:s}
\end{figure}

Fig.~\ref{fig:P_BL} and Fig.~\ref{fig:P} plot the error rates of the baseline schemes and the proposed solutions versus the number of channel taps $P$, respectively.
From Fig.~\ref{fig:P_BL} and Fig.~\ref{fig:P}, we can see that the error rate of each scheme increases with $P$. The error rates of   {\em BL-AMP-act.} and {\em BL-ML-act.} rapidly increase with $P$, as the statistical difference between flat Rayleigh fading and frequency selective Rayleigh fading becomes more significant at larger $P$. For each activity detection scheme for virtual devices, the increase of the error rate with $P$ mainly derives from the increase of the number of the virtual devices, $NP$, with $P$.
The slight increases of the error rates of {\em Prop.-ML-act.} and {\em Prop.-MAP-act.} with $P$ when $P\geq 3$   are mainly due to the numerical error for numerically calculating the roots of polynomials with degrees $2p-1$ and $2p$, respectively. By comparing Fig.~\ref{fig:P_BL} and Fig.~\ref{fig:P}, we see that the proposed solutions significantly outperform the baseline schemes, especially at large $P$. This is why we plot their error rates in separate figures. As the baseline schemes yield much higher error rates, we do not evaluate them in Fig.~\ref{fig:L}, Fig.~\ref{fig:M}, Fig.~\ref{fig:K}, and Fig.~\ref{fig:s}.

Fig.~\ref{fig:L}, Fig.~\ref{fig:M}, and Fig.~\ref{fig:K} plot the error rates of the proposed solutions versus the pilot length $L$, number of antennas $M$, and activity probability $q$
in the i.i.d. case and correlated case, respectively. From Fig.~\ref{fig:L} and Fig.~\ref{fig:M}, we can observe that the error rate of each proposed solution decreases with $L$ and $M$, mainly due to more measurement vectors adopted and more observations available, respectively. From Fig.~\ref{fig:K}, we can see that the error rate of each proposed solution increases with $q$, mainly due to the increase in the number of active devices on average. Fig.~\ref{fig:s} shows the error rates of the proposed solutions versus the group size $\frac{N}{K}$ in the correlated case.  From Fig.~\ref{fig:s}, we have the following observations. The error rate of each proposed ML estimation-based solution increases with $\frac{N}{K}$, due to the increment of the variance of the number of active devices and the reduction of the sample space of device activities. In contrast, each proposed MAP estimation-based solution decreases with $\frac{N}{K}$, due to proper exploitation of the correlation among devices.

We can make the following observations from Fig.~\ref{fig:L}, Fig.~\ref{fig:M}, Fig.~\ref{fig:K}, and Fig.~\ref{fig:s}. Firstly, {\em Prop.-ML-virt.} and {\em Prop.-MAP-virt.} outperform {\em Prop.-ML-virt.-rel.} and {\em Prop.-MAP-virt.-rel.}, respectively, in the i.i.d. and correlated cases, as {\em Prop.-ML-virt.} and {\em Prop.-MAP-virt.} address the issue of the coupling constraints for the $NP$ virtual devices. Secondly, {\em Prop.-ML-act.}  and {\em Prop.-MAP-act.} achieve lower error rates than  {\em Prop.-ML-virt.} and {\em Prop.-MAP-virt.}, respectively, in the i.i.d. and correlated cases, as a problem with a smaller size can be more effectively solved. Thirdly, the proposed MAP estimation-based solutions outperform the corresponding ML estimation-based solutions due to the utilization of the prior distribution of device activities.
Note that the relative gain of each proposed MAP estimation-based solution over its ML counterpart in the i.i.d. case is smaller than that in the correlated case, as the correlation among device activities has a considerable impact on device activity detection. Besides, note that in each case, such gain  decreases with $M$ and $q$, since the influence of the prior distribution of device activities reduces as the number of observations increases (as discussed in Section~\ref{sec:map}) and the difference between the active probability $q$ and inactive probability $1-q$ decreases. As a result, each proposed MAP-estimation-based solution is more preferable than its ML counterpart at small $M$, $q$, and $\frac{N}{K}$ provided that the prior distribution of device activities is available.

%

\begin{figure}[t]
\begin{center}
\subfigure[\small{Length of channel taps $P$ at $L=48$ and $M=256$.}\label{fig:error_vs_time1}]
{\resizebox{5cm}{!}{\includegraphics{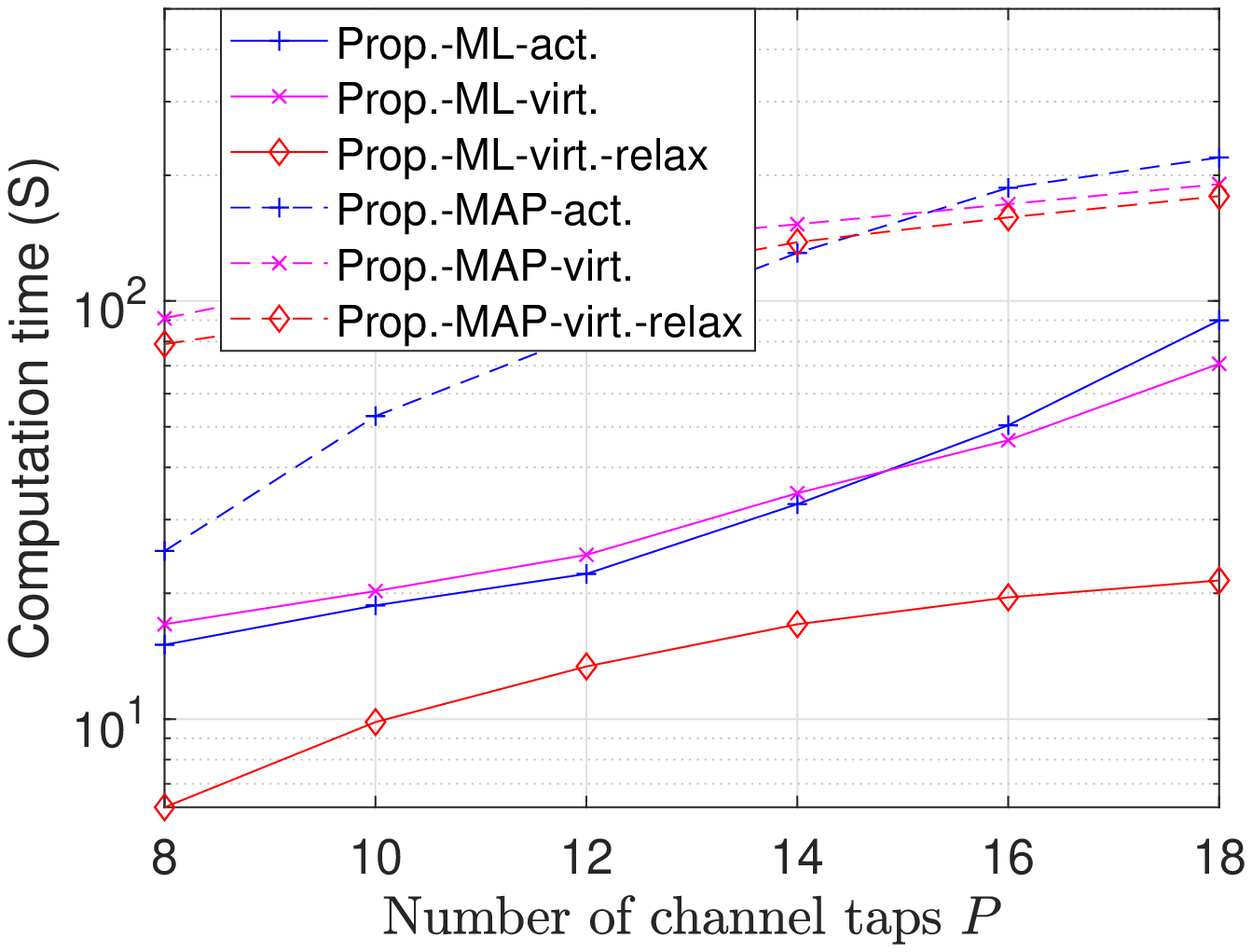}}}\quad
\subfigure[\small{Length of channel taps $P$ at $L=32$ and $M=256$.}\label{fig:error_vs_time2}]
{\resizebox{5cm}{!}{\includegraphics{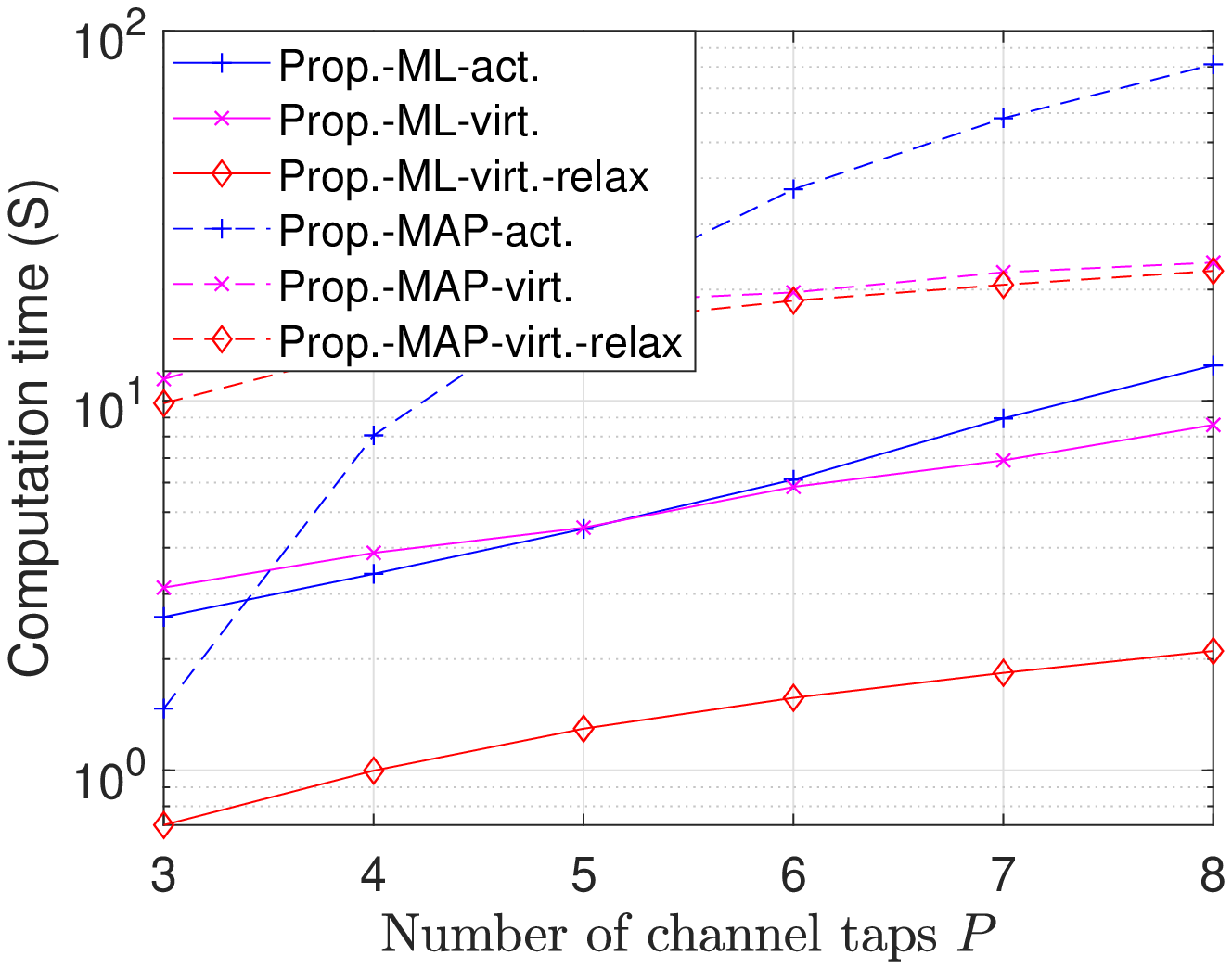}}}
\end{center}
\vspace{-4mm}
\caption{\small{Computation times of proposed solutions versus number of channel taps $P$.}}
\vspace{-9mm}
\label{fig:rate_s}
\end{figure}

Fig.~\ref{fig:error_vs_time1} and Fig.~\ref{fig:error_vs_time2} plot the computation times versus the number of channel taps $P$ at different pilot lengths.
We have the following observations from these figures. Firstly, {\em Prop.-ML-virt.-rel.} and {\em Prop.-MAP-virt.-rel.} have shorter  computation times than {\em Prop.-ML-virt.} and {\em Prop.-MAP-virt.}, respectively, as they do not consider the coupling constraints for the $NP$ virtual devices. Secondly, {\em Prop.-ML-act.} and {\em Prop.-MAP-act.} have  shorter  computation times than {\em Prop.-ML-virt.}, {\em Prop.-ML-virt.-rel.} and {\em Prop.-MAP-virt.}, {\em Prop.-MAP-virt.-rel.} at small $P$, respectively, as each corresponding problem has a smaller number of variables, and the computational complexity for numerically calculating the roots of a polynomial with a smaller degree is only slightly larger than that for analytically calculating the roots of a polynomial with degree $3$.
Thirdly, {\em Prop.-ML-act.} and {\em Prop.-MAP-act.} have  longer  computation times than {\em Prop.-ML-virt.}, {\em Prop.-ML-virt.-rel.} and {\em Prop.-MAP-virt.}, {\em Prop.-MAP-virt.-rel.} at large $P$, respectively, as the computational complexity for numerically calculating the roots of a polynomial with a larger degree is significantly larger than that for analytically calculating the roots of a polynomial with degree $3$. Fourthly, each proposed MAP-estimation-based solution has a longer computation time than its ML counterpart, as the incorporating of the prior distribution of device activities yields a more complex problem. Last but not least, for the three proposed ML-based or MAP-based solutions, the preferable regime in terms of computation time hinges on the value of $L$.


\vspace{-4mm}
\section{conclusion}
This paper investigated statistical device activity detection for massive grant-free access in a wideband system under frequency-selective fading. Specifically, we presented an OFDM-based massive grant-free access scheme and obtained two models for the received pilot signal, which are critical for designing various device activity detection and channel estimation methods for OFDM-based massive grant-free access. Moreover,  we proposed three ML-based estimation methods and three MAP-based estimation methods, respectively, for device activity detection under frequency-selective Rayleigh fading. Finally, analytical and numerical results show that the proposed statistical device activity detection methods achieve different detection accuracies with different computation times and can meet diverse practical needs for OFDM-based massive grant-free access.
\vspace{-3mm}
\section*{Appendix A: Proof of Theorem~\ref{Thm:Step_APs_Penal_M_non_extension}}
For notation simplicity, we denote $\boldsymbol\Gamma_{\boldsymbol\alpha,n}\triangleq\mathbf S_{n}^H\mathbf \Sigma^{(1)-1}_{\boldsymbol\alpha}\mathbf S_{n}$ and $\widehat{\boldsymbol\Gamma}_{\boldsymbol\alpha,n} \triangleq \mathbf{S}_n^H\boldsymbol\Sigma_{\boldsymbol\alpha}^{(1)-1}
\widehat{\mathbf \Sigma}_{\mathbf R}\boldsymbol\Sigma_{\boldsymbol\alpha}^{(1)-1}\mathbf{S}_n$ in Appendix A. Before proving Theorem~\ref{Thm:Step_APs_Penal_M_non_extension}, we first show some preliminary results.
By the eigenvalue decomposition of  $\boldsymbol\Gamma_{\boldsymbol\alpha,n}$ given in \eqref{eq:eigenvalue_decomposition}, we have the following results:
\begin{align}
 & (\mathbf{I}_P+dg_n\boldsymbol\Gamma_{\boldsymbol\alpha,n})^{-1}
      = \mathbf{U}_n(\mathbf{I}_P+dg_n{\rm diag}(\mathbf{v}))^{-1}\mathbf{U}_n^H \label{eq:middle}\\
&   =
   \mathbf{U}_n{\rm diag}\left(\left[\frac{1}{1+v_1g_nd},...,\frac{1}{1+v_pg_nd}\right]\right)\mathbf{U}_n^H, \label{eq:eigen_I}
\\
&  \text{tr}\left(   (\mathbf{I}_P+dg_n\boldsymbol\Gamma_{\boldsymbol\alpha,n})^{-1}
    \boldsymbol\Gamma_{\boldsymbol\alpha,n}\right) \eqla
    \text{tr}\left(   \mathbf{U}_n{\rm diag}\left(\left[\frac{1}{1+v_1g_nd},...,\frac{1}{1+v_pg_nd}\right]\right)\mathbf{U}_n^H
    \mathbf{U}_n{\rm diag}\left(\mathbf{v}\right)\mathbf{U}_n^H\right) \notag
    \\& \eqlb  \text{tr}\left(   \mathbf{U}_n{\rm diag}\left(\left[\frac{v_1}{1+v_1g_nd},...,\frac{v_p}{1+v_pg_nd}\right]\right)\mathbf{U}_n^H\right)
 \eqlc \sum_{p\in\mathcal{P}}\frac{v_p}{1+v_pg_nd},\label{eq:same_evector}
\\
& \text{tr}\left( (\mathbf{I}_P+dg_n\boldsymbol\Gamma_{\boldsymbol\alpha,n})^{-2}
  \widehat{\boldsymbol\Gamma}_{\boldsymbol\alpha,n}\right) =\text{tr}\left({\mathbf{U}_n\rm diag}\left(\left[\frac{1}{(1+v_1g_nd)^2},...,\frac{1}{(1+v_pg_nd)^2}\right]\right) \mathbf{U}_n^H \widehat{\boldsymbol\Gamma}_{\boldsymbol\alpha,n} \right)
\notag  \\& \eqld
  \text{tr}\left({\rm diag}\left(\left[\frac{1}{(1+v_1g_nd)^2},...,\frac{1}{(1+v_pg_nd)^2}\right]\right) \mathbf{U}_n^H \widehat{\boldsymbol\Gamma}_{\boldsymbol\alpha,n}\mathbf{U}_n \right)
  =\sum_{p\in\mathcal{P}}\frac{u_p}{(1+v_pg_nd)^2}, \label{eq:cycle}
\\
 & \lvert\boldsymbol\Sigma_{\boldsymbol\alpha}^{(1)}+dg_n\mathbf{S}_n\mathbf{S}_n^H\rvert = \lvert\boldsymbol\Sigma_{\boldsymbol\alpha}^{(1)}\rvert \lvert \mathbf{I}_{L}+dg_n\boldsymbol\Sigma_{\boldsymbol\alpha}^{(1)-1}\mathbf{S}_n\mathbf{S}_n^H\rvert \eqle \lvert\boldsymbol\Sigma_{\boldsymbol\alpha}^{(1)}\rvert \lvert \mathbf{I}_{P}+dg_n\mathbf{S}_n^H\boldsymbol\Sigma_{\boldsymbol\alpha}^{(1)-1}\mathbf{S}_n\rvert, \label{eq:Sylvester_determinant_identity} \\
&(\boldsymbol\Sigma_{\boldsymbol\alpha}^{(1)}+dg_n\mathbf{S}_n\mathbf{S}_n^H)^{-1} \eqlf \mathbf \Sigma_{\boldsymbol\alpha}^{(1)-1}-dg_n\boldsymbol\Sigma_{\boldsymbol\alpha}^{(1)-1}
\mathbf{S}_n(\mathbf{I}_P+dg_n\boldsymbol\Gamma_{\boldsymbol\alpha,n})^{-1}\mathbf{S}_n^H\boldsymbol\Sigma_{\boldsymbol\alpha}^{-1}.  \label{eq:Sherman}
\end{align}
Here, $(a)$ is due to the eigenvalue decompositions of $\boldsymbol\Gamma_{\boldsymbol\alpha,n}$ and $(\mathbf{I}_P+dg_n\boldsymbol\Gamma_{\boldsymbol\alpha,n})^{-1}$ in \eqref{eq:eigenvalue_decomposition} and \eqref{eq:middle}; $(b)$ is due to $\mathbf{U}_n^H
    \mathbf{U}_n = \mathbf{I}_P$; $(c)$ is due to the fact that the trace of a matrix equals to the sum of the eigenvalues of the matrix; $(d)$ is due to  the cyclic property
of trace; $(e)$  is due to Sylvester's determinant identity for any $\boldsymbol \Sigma^{(1)}_{\boldsymbol\alpha}\succ 0$, and $(f)$ is due to Sherman-Morrison-Woodbury formula.

In what follows, we show Theorem~\ref{Thm:Step_APs_Penal_M_non_extension} based on \eqref{eq:eigen_I}-\eqref{eq:Sherman}.
First, we have:
\begin{align}
& f^{(1)}(\boldsymbol\alpha+d\mathbf{e}_n) \eqlg   \log\lvert\boldsymbol\Sigma_{\boldsymbol\alpha}^{(1)}+dg_n\mathbf{S}_n\mathbf{S}_n^H\rvert
  +\text{tr}((\boldsymbol\Sigma_{\boldsymbol\alpha}^{(1)}+dg_n\mathbf{S}_n\mathbf{S}_n^H)^{-1}\widehat{\mathbf \Sigma}_{\mathbf R}) \notag \\
& \eqlh  \log\left(|\mathbf \Sigma_{\boldsymbol\alpha}^{(1)}|\lvert\mathbf{I}_P +dg_{n}\boldsymbol\Gamma_{\boldsymbol\alpha,n}\rvert\right)+  \text{tr}\left(\left(\mathbf \Sigma_{\boldsymbol\alpha}^{(1)-1}-dg_n\boldsymbol\Sigma_{\boldsymbol\alpha}^{(1)-1}
\mathbf{S}_n(\mathbf{I}_P+dg_n\boldsymbol\Gamma_{\boldsymbol\alpha,n})^{-1}\mathbf{S}_n^H\boldsymbol\Sigma_{\boldsymbol\alpha}^{-1}\right)\widehat{\mathbf \Sigma}_{\mathbf R} \right)\notag\\
& \eqli f^{(1)}(\boldsymbol\alpha)+f^{(1)}_{\boldsymbol\alpha,n}(d),\label{eqn:ML_expansion_a}
\end{align}
where $(g)$ is due to \eqref{eqn:f_ml}, $(h)$ is due to \eqref{eq:Sylvester_determinant_identity} and \eqref{eq:Sherman}, and $(i)$ is due to the cyclic property of trace.  Next, we have:
\begin{align}
  &  \frac{\partial f^{(1)}(\boldsymbol\alpha+d\mathbf{e}_n)}{\partial d} \eqlj \frac{\partial f^{(1)}_{\boldsymbol\alpha,n}(d)}{\partial d} \eqlk
 g_n\left( \text{tr}\left(\left(\mathbf{I}_P+dg_n\boldsymbol\Gamma_{\boldsymbol\alpha,n}\right)^{-1}\boldsymbol\Gamma_{\boldsymbol\alpha,n}\right)
  -\text{tr}\left(\left(\mathbf{I}_P+dg_n\boldsymbol\Gamma_{\boldsymbol\alpha,n}\right)^{-2}
  \widehat{\boldsymbol\Gamma}_{\boldsymbol\alpha,n}\right)\right) \nonumber \\
&  \eqll  g_n\sum_{p\in\mathcal{P}}\bigg(\frac{v_p}{1+v_pg_nd}-\frac{u_p}{(1+v_pg_nd)^2}\bigg)
=\frac{g^{(1)}_{\boldsymbol\alpha,n}(d)}{\prod_{p\in\mathcal{P}}(1+v_pg_nd)^2},\label{eq:derivation_f1}
\end{align}
where $(j)$ is due to \eqref{eqn:ML_expansion_a}, $(k)$ is due to \eqref{eq:function}, $(l)$ is due to  \eqref{eq:same_evector} and \eqref{eq:cycle}.
Thus, the solution of $  \frac{\partial f^{(1)}(\boldsymbol\alpha+d\mathbf{e}_n)}{\partial d}=0$, i.e., the solution of $g^{(1)}_{\boldsymbol\alpha,n}(d)=0$,
is given by $\mathcal{D}_n^{(1)}$. Therefore, combining with $\alpha_n\in [0,1],n\in\mathcal{N}$,  we can obtain the optimal solution of \eqref{eqn:Penal_a_ML_extension} as in \eqref{eqn:d_Penalty_a_extension}.
\vspace{-4mm}
\section*{Appendix B: Proof of Theorem~\ref{Thm:Step_APs_Penal_M}}
Following the proofs for \eqref{eq:Sylvester_determinant_identity} and \eqref{eq:Sherman}, we can have:
\begin{align}
(\mathbf \Sigma_{\boldsymbol\beta}+d\delta_i\mathbf S_{:,i}\mathbf S_{:,i}^H)^{-1}=& \mathbf \Sigma^{-1}_{\boldsymbol\beta}-\frac{d\delta_i\mathbf \Sigma^{-1}_{\boldsymbol\beta}\mathbf{S}_{:,i}\mathbf{S}_{:,i}^H\mathbf \Sigma^{-1}_{\boldsymbol\beta}}{1+d\delta_i\mathbf{S}_{:,i}^H\mathbf \Sigma^{-1}_{\boldsymbol\beta}\mathbf{S}_{:,i}}, \label{eq:penal1}
  \\
\lvert\boldsymbol\Sigma_{\boldsymbol\beta}^{(2)}+d\delta_i\mathbf{S}_{:,i}\mathbf{S}_{:,i}^H\rvert = &
\lvert\boldsymbol\Sigma_{\boldsymbol\beta}^{(2)} \rvert(1+d\delta_i\mathbf S_{:,i}^H\mathbf \Sigma_{\boldsymbol\beta}^{(2)-1}\mathbf S_{:,i}). \label{eq:penal2}
\end{align}
Now, we show Theorem~\ref{Thm:Step_APs_Penal_M}. First, we have:
\begin{align}
 & \tilde{f}^{(2)}(\boldsymbol \beta+d\mathbf{e}_i) \eqla   \log\lvert\boldsymbol\Sigma_{\boldsymbol\beta}^{(2)}+d\delta_i\mathbf{S}_{:,i}\mathbf{S}_{:,i}^H\rvert
  +\text{tr}((\boldsymbol\Sigma_{\boldsymbol\beta}^{(2)}+d\delta_i\mathbf{S}_{:,i}\mathbf{S}_{:,i}^H)^{-1}\widehat{\mathbf \Sigma}_{\mathbf R}) +\rho\eta(\boldsymbol\beta+d\mathbf{e}_i) \notag \\
 &\eqlb \log\left(|\mathbf \Sigma_{\boldsymbol\beta}^{(2)}|(1+d\delta_i\mathbf S_{:,i}^H\mathbf \Sigma_{\boldsymbol\beta}^{(2)-1}\mathbf S_{:,i})\right)+\text{tr}\left(\left(\mathbf \Sigma_{\boldsymbol\beta}^{(2)-1}-\frac{d\delta_i\mathbf \Sigma_{\boldsymbol\beta}^{(2)-1}\mathbf S_{:,i}\mathbf S_{:,i}^H \mathbf \Sigma_{\boldsymbol\beta}^{(2)-1}}{1+d\delta_i\mathbf S_{:,i}^H\mathbf \Sigma_{\boldsymbol\beta}^{(2)-1}\mathbf S_{:,i}}\right)\widehat{\mathbf \Sigma}_{\mathbf R} \right) +\rho\eta(\boldsymbol\beta+d\mathbf{e}_i)  \notag\\
 &  \eqlc \tilde{f}^{(2)}(\boldsymbol\beta)+\tilde{f}^{(2)}_{\boldsymbol\beta,i}(d), \label{eqn:ML_expansion_a_extension}
\end{align}
where $(a)$ is due to \eqref{eqn:f_ml_ex}, $(b)$ is due to \eqref{eq:penal1} and \eqref{eq:penal2}, and $(c)$ is due to the cyclic property of trace. Then, we have:
\begin{align}
  &\frac{\partial  \tilde{f}^{(2)}(\boldsymbol\beta+d\mathbf{e}_i)}{\partial d} \eqld \frac{\partial \tilde{f}^{(2)}_{\boldsymbol\beta,i}(d)}{\partial d} \eqle \frac{\delta_i\mathbf{S}_{:,i}^H\mathbf \Sigma^{-1}_{\boldsymbol\beta}\mathbf{S}_{:,i}(1+d\delta_i\mathbf{S}_{:,i}^H\mathbf \Sigma^{-1}_{\boldsymbol\beta}\mathbf{S}_{:,i})-\delta_i\mathbf S_{:,i}^H\mathbf \Sigma^{(2)-1}_{\boldsymbol\beta}\widehat{\mathbf \Sigma}_{\mathbf R}\mathbf \Sigma^{(2)-1}_{\boldsymbol\beta}\mathbf S_{:,i}}{(1+d\delta_i\mathbf{S}_{:,i}^H\mathbf \Sigma^{-1}_{\boldsymbol\beta}\mathbf{S}_{:,i})^2} \nonumber \\ & +\frac{\rho}{P}\bigg(1-\frac{2}{P}\sum\limits_{p=1}^{P}\beta_{\left(\lceil \frac{i}{P} \rceil-1\right)P+p}\bigg)=\frac{\tilde{g}^{(2)}_{\boldsymbol\beta,i}(d)}{(1+d\delta_i\mathbf{S}_{:,i}^H\mathbf \Sigma^{-1}_{\boldsymbol\beta}\mathbf{S}_{:,i})^2},\label{eq:appendixB_d}
\end{align}
where $(d)$ is given by \eqref{eqn:ML_expansion_a_extension}, and $(e)$ follows the derivation of (22) in \cite{Caire18ISIT}.
Thus, the solution of $ \frac{\partial  \tilde{f}^{(2)}(\boldsymbol\beta+d\mathbf{e}_i)}{\partial d}=0$, i.e., the solution of $\tilde{g}^{(2)}_{\boldsymbol\beta,i}(d)=0$, is given by $\mathcal{D}_i^{(2)}$. Therefore, combining with $\beta_i\in[0,1],i\in\mathcal{I}$, we can obtain the optimal solution of
\eqref{eqn:Penal_a} as in \eqref{eqn:d_Penalty_a}.
\vspace{-3mm}
\section*{Appendix C: Proof of Theorem~\ref{Thm:Step_APs_Penal_M_non_extension_map}}
First, we have:
\begin{align}
 & f^{(3)}(\boldsymbol\alpha+d\mathbf{e}_n)
\eqla f^{(1)}(\boldsymbol\alpha+d\mathbf{e}_n)-\frac{1}{M}\sum_{\omega\in\Psi}\bigg(c_{\alpha,\omega}\prod_{n\in\omega}
(\alpha_n+d\mathbf e_n)\bigg)
\eqlb  f^{(1)}(\boldsymbol\alpha)+f^{(1)}_{\boldsymbol\alpha,n}(d) \nonumber \\ &- \frac{1}{M}\bigg(\sum_{\omega\in\Psi}\bigg(c_{\omega}\prod_{n\in\omega}\alpha_n)\bigg)+
d\sum_{\omega\in\Psi:n\in\omega}\bigg(c_{\omega}\prod_{n^{'}\in\omega,n^{'}\neq n}\alpha_{n'}\bigg)\bigg) = f^{(3)}(\boldsymbol\alpha)+f^{(3)}_{\boldsymbol\alpha,n}(d),\label{eqn:ML_expansion_a_map1}
\end{align}
where $(a)$ is due to \eqref{eqn:f_map}, and $(b)$ is due to \eqref{eqn:ML_expansion_a}.
Next, we have:
\begin{align*}
 & \frac{\partial f^{(3)}(\boldsymbol\alpha+d\mathbf{e}_n)}{\partial d} \eqlc \frac{\partial f^{(3)}_{\boldsymbol\alpha,n}(d)}{\partial d}   \eqld \frac{\partial f^{(1)}_{\boldsymbol\alpha,n}(d)}{\partial d}  - \frac{\epsilon_n(\boldsymbol\alpha)}{M}  \eqle \frac{g^{(3)}_{\boldsymbol\alpha,n}(d)}{\prod_{p\in\mathcal{P}}
  (1+v_pdg_n)^2},
\end{align*}
where $(c)$ is due to \eqref{eqn:ML_expansion_a_map1}, $(d)$ is based on the expression of \eqref{eq:function_map}, and $(e)$ is due to \eqref{eq:derivation_f1}.
Thus, the solution of $\frac{\partial f^{(3)}(\boldsymbol\alpha+d\mathbf{e}_n)}{\partial d}=0$, i.e., the solution of $g^{(3)}_{\boldsymbol\alpha,n}(d)=0$,  is given by $\mathcal{D}_n^{(3)}$. Therefore, combining with $\alpha_n\in[0,1],n\in\mathcal{N}$, we can obtain the optimal solution of \eqref{eqn:Penal_a_MAP_extension} as in \eqref{eqn:d_Penalty_a_extension_map}.
\vspace{-3mm}
\section*{Appendix D: Proof of Theorem~\ref{Thm:Step_APs_Penal_M1}}
First, we have:
\begin{align}
 & \tilde{f}^{(4)}(\boldsymbol\beta+d\mathbf{e}_i) \eqla \tilde{f}^{(2)}(\boldsymbol\beta+d\mathbf{e}_i)- \frac{1}{M}\sum_{\omega\in\Psi}\bigg(c_{\omega}\prod_{n\in\omega}\frac{\sum\limits_{p\in\mathcal{P}}
\beta_{(n-1)P+p}+d\mathbf{e}_i}{P}\bigg) \eqlb  \tilde{f}^{(2)}(\boldsymbol\beta)+f^{(2)}_{\boldsymbol\beta,i}(d) \nonumber \\ & -  \frac{1}{M}\sum_{\omega\in\Psi}\bigg(c_{\omega}\prod_{n\in\omega}\frac{\sum\limits_{p\in\mathcal{P}}
\beta_{(n-1)P+p}}{P}\bigg)- \frac{d}{M}\sum_{\omega\in\Psi:i\in\omega}\bigg(c_{\omega}\prod_{i^{'}\in\omega,i^{'}\neq i}\frac{\sum\limits_{p\in\mathcal{P}}\beta_{(i'-1)P+p}}{P}\bigg)  = \tilde{f}^{(4)}(\boldsymbol\beta)+f^{(4)}_{\boldsymbol\beta,i}(d), \label{eqn:ML_expansion_a_extension_M1}
\end{align}
where $(a)$ is due to \eqref{eqn:f_map}, and $(b)$ is due to \eqref{eqn:ML_expansion_a_extension}.
Next, we have:
\begin{align*}
 & \frac{\partial \tilde{f}^{(4)}(\boldsymbol\beta+d\mathbf{e}_i)}{\partial d} \eqlc \frac{\partial \tilde{f}^{(4)}_{\boldsymbol\beta,i}(d)}{\partial d}  \eqld \frac{\partial \tilde{f}^{(2)}_{\boldsymbol\beta,i}(d)}{\partial d} -\frac{1}{M}\sum_{\omega\in\Psi:i\in\omega}\bigg(c_{\omega}\prod_{i^{'}\in\omega,i^{'}\neq i}\frac{\sum\limits_{p\in\mathcal{P}}\beta_{(i'-1)P+p}}{P}\bigg)
 \\ & \eqle \frac{\tilde{g}^{(2)}_{\boldsymbol\beta,i}(d)}{(1+d\delta_i\mathbf{S}_{:,i}^H\mathbf \Sigma^{-1}_{\boldsymbol\beta}\mathbf{S}_{:,i})^2} -\frac{1}{M}\sum_{\omega\in\Psi:i\in\omega}\bigg(c_{\omega}\prod_{i^{'}\in\omega,i^{'}\neq i}\frac{\sum\limits_{p\in\mathcal{P}}\beta_{(i'-1)P+p}}{P}\bigg)  =  \frac{\tilde{g}^{(4)}_{\boldsymbol\beta,i}(d)}{(1+d\delta_i\mathbf{S}_{:,i}^H\mathbf \Sigma^{-1}_{\boldsymbol\beta}\mathbf{S}_{:,i})^2},
\end{align*}
where $(c)$ is due to \eqref{eqn:ML_expansion_a_extension_M1}, $(d)$ is based on the expression of \eqref{eq:three_times}, and $(e)$ is due to \eqref{eq:appendixB_d}.
Thus, the optimal solution of $\frac{\partial \tilde{f}^{(4)}(\boldsymbol\beta+d\mathbf{e}_i)}{\partial d}=0$, i.e., the solution of $\tilde{g}^{(4)}_{\boldsymbol\beta,i}(d)=0$, is given by $\mathcal{D}_i^{(4)}$. Therefore, combining with $\beta_i\in[0,1],i\in\mathcal{I}$, we can obtain the optimal solution of \eqref{eqn:Penal_MAP_a} as in \eqref{eqn:d_Penalty_a_MAP}.

\bibliographystyle{IEEEtran}

\begin{thebibliography}{10}
\providecommand{\url}[1]{#1}
\csname url@samestyle\endcsname
\providecommand{\newblock}{\relax}
\providecommand{\bibinfo}[2]{#2}
\providecommand{\BIBentrySTDinterwordspacing}{\spaceskip=0pt\relax}
\providecommand{\BIBentryALTinterwordstretchfactor}{4}
\providecommand{\BIBentryALTinterwordspacing}{\spaceskip=\fontdimen2\font plus
\BIBentryALTinterwordstretchfactor\fontdimen3\font minus
  \fontdimen4\font\relax}
\providecommand{\BIBforeignlanguage}[2]{{%
\expandafter\ifx\csname l@#1\endcsname\relax
\typeout{** WARNING: IEEEtran.bst: No hyphenation pattern has been}%
\typeout{** loaded for the language `#1'. Using the pattern for}%
\typeout{** the default language instead.}%
\else
\language=\csname l@#1\endcsname
\fi
#2}}
\providecommand{\BIBdecl}{\relax}
\BIBdecl

\bibitem{YuhangGLOBECOM}
Y.~{Jia}, Y.~{Cui}, and W.~{Jiang}, ``{Device} activity detection for
  grant-free massive access under frequency-selective {Rayleigh} fading,'' in
  \emph{Proc. IEEE GLOBECOM}, Dec. 2021.

\bibitem{Erik}
L.~Liu, E.~G. Larsson, W.~Yu, P.~Popovski, C.~Stefanovic, and E.~de~Carvalho,
  ``{Sparse} signal processing for grant-free massive connectivity: A future
  paradigm for random access protocols in the internet of things,'' \emph{IEEE
  Signal Process. Mag.}, vol.~35, no.~5, pp. 88--99, Sept. 2018.

\bibitem{Chen18TSP}
Z.~{Chen}, F.~{Sohrabi}, and W.~{Yu}, ``{Sparse} activity detection for massive
  connectivity,'' \emph{IEEE Trans. Signal Process.}, vol.~66, no.~7, pp.
  1890--1904, Apr. 2018.

\bibitem{Liu18TSP}
L.~{Liu} and W.~{Yu}, ``{Massive} connectivity with massive {MIMO}-part {I}:
  Device activity detection and channel estimation,'' \emph{IEEE Trans. Signal
  Process.}, vol.~66, no.~11, pp. 2933--2946, Jun. 2018.

\bibitem{Senel18TCOM}
K.~{Senel} and E.~G. {Larsson}, ``{Grant}-free massive {MTC}-enabled massive
  {MIMO}: A compressive sensing approach,'' \emph{IEEE Trans. Commun.},
  vol.~66, no.~12, pp. 6164--6175, Dec. 2018.

\bibitem{Shao19IoTJ}
X.~{Shao}, X.~{Chen}, C.~{Zhong}, J.~{Zhao}, and Z.~{Zhang}, ``A unified design
  of massive access for cellular internet of things,'' \emph{IEEE Internet of
  Things J.}, vol.~6, no.~2, pp. 3934--3947, Apr. 2019.

\bibitem{Chen19TWC}
Z.~{Chen}, F.~{Sohrabi}, and W.~{Yu}, ``{Multi}-cell sparse activity detection
  for massive random access: massive {MIMO} versus cooperative {MIMO},''
  \emph{IEEE Trans. Wireless Commun.}, vol.~18, no.~8, pp. 4060--4074, Aug.
  2019.

\bibitem{8536396}
T.~Jiang, Y.~Shi, J.~Zhang, and K.~B. Letaief, ``{Joint} activity detection and
  channel estimation for {IoT} networks,'' \emph{IEEE Internet of Things J.}

\bibitem{JSAC_li}
Y.~Cui, S.~Li, and W.~Zhang, ``{Jointly} sparse signal recovery and support
  recovery via deep learning with applications in {MIMO}-based grant-free
  random access,'' \emph{IEEE J. Sel. Areas Commun.}, vol.~39, no.~3, pp.
  788--803, Mar. 2021.

\bibitem{9413733}
H.~Djelouat, M.~Leinonen, and M.~Juntti, ``{Iterative} reweighted algorithms
  for joint user identification and channel estimation in spatially correlated
  massive {MTC},'' in \emph{Proc. IEEE ICASSP}, Jun. 2021, pp. 4805--4809.

\bibitem{9522070}
T.~Li, J.~Zhang, Z.~Yang, Z.~L. Yu, Z.~Gu, and Y.~Li, ``{Dynamic} user activity
  and data detection for grant-free {NOMA} via weighted $\ell_{2,1}$
  minimization,'' \emph{to appear in IEEE Trans. Wireless Commun.}, 2021.

\bibitem{Caire18ISIT}
A.~{Fengler}, S.~{Haghighatshoar}, P.~{Jung}, and G.~{Caire}, ``{Non}-bayesian
  activity detection, large-scale fading coefficient estimation, and unsourced
  random access with a massive {MIMO} receiver,'' \emph{IEEE Trans. Inf.
  Theory}, vol.~67, no.~5, pp. 2925--2951, May 2021.

\bibitem{Yu19ICC}
Z.~{Chen}, F.~{Sohrabi}, Y.~{Liu}, and W.~{Yu}, ``{Covariance} based joint
  activity and data detection for massive random access with massive {MIMO},''
  in \emph{Proc. IEEE ICC}, May 2019, pp. 1--6.

\bibitem{zhongGLOBECOM}
X.~Shao, X.~Chen, D.~W.~K. Ng, C.~Zhong, and Z.~Zhang, ``{Covariance}-based
  cooperative activity detection for massive grant-free random access,'' in
  \emph{Proc. IEEE GLOBECOM}, Dec. 2020, pp. 1--6.

\bibitem{Yu21SPAWC}
Z.~{Wang}, Y.~{Liu}, Z.~{Chen}, and W.~{Yu}, ``{Accelerating} coordinate
  descent via active set selection for device activity detection for multi-cell
  massive random access,'' in \emph{Proc. IEEE SPAWC}, Sept. 2021, pp. 1--5.

\bibitem{JiangSPAWC}
D.~Jiang and Y.~Cui, ``{MAP}-based pilot state detection in grant-free random
  access for {mMTC},'' in \emph{Proc. IEEE SPAWC}, May 2020, pp. 1--5.

\bibitem{Jiang21TWC}
D.~{Jiang} and Y.~{Cui}, ``{ML} and {MAP} device activity detections for
  grant-free massive access in multi-cell networks,'' \emph{to appear in IEEE
  Trans. Wireless Commun.}, 2021.

\bibitem{LiSPL}
S.~Li, W.~Zhang, Y.~Cui, H.~V. Cheng, and W.~Yu, ``{Joint} design of
  measurement matrix and sparse support recovery method via deep
  auto-encoder,'' \emph{IEEE Signal Process. Lett.}, vol.~26, no.~12, pp.
  1778--1782, Dec. 2019.

\bibitem{ZhangSPAWC}
W.~Zhang, S.~Li, and Y.~Cui, ``{Jointly} sparse support recovery via deep
  auto-encoder with applications in {MIMO}-based grant-free random access for
  {mMTC},'' in \emph{Proc. IEEE SPAWC}, May 2020, pp. 1--5.

\bibitem{LiWCNC}
S.~Li, W.~Zhang, and Y.~Cui, ``{Jointly} sparse signal recovery via deep
  auto-encoder and parallel coordinate descent unrolling,'' in \emph{Proc. IEEE
  WCNC}, May 2020, pp. 1--6.

\bibitem{9484069}
Z.~Mao, X.~Liu, M.~Peng, Z.~Chen, and G.~Wei, ``{Joint} channel estimation and
  active-user detection for massive access in internet of things – a deep
  learning approach,'' \emph{to appear in IEEE Internet of Things J.}, 2021.

\bibitem{9508782}
Y.~Shi, H.~Choi, Y.~Shi, and Y.~Zhou, ``{Algorithm} unrolling for massive
  access via deep neural network with theoretical guarantee,'' \emph{to appear
  in IEEE Trans. Wireless Commun.}, Aug. 2021.

\bibitem{9146533}
Y.~Qiang, X.~Shao, and X.~Chen, ``{A} model-driven deep learning algorithm for
  joint activity detection and channel estimation,'' \emph{IEEE Commun. Lett.},
  vol.~24, no.~11, pp. 2508--2512, Jul. 2020.

\bibitem{9432908}
X.~Shao, X.~Chen, Y.~Qiang, C.~Zhong, and Z.~Zhang, ``{Feature}-aided
  adaptive-tuning deep learning for massive device detection,'' \emph{IEEE J.
  Sel. Commun.}, vol.~39, no.~7, pp. 1899--1914, Jul. 2021.

\bibitem{dahlman20205g}
E.~Dahlman, S.~Parkvall, and J.~Skold, \emph{5{G} {NR}: {The} next generation
  wireless access technology}.\hskip 1em plus 0.5em minus 0.4em\relax Academic
  Press, 2020.

\bibitem{9296241}
T.~Hara, H.~Iimori, and K.~Ishibashi, ``{Hyperparameter}-free receiver for
  grant-free {NOMA} systems with {MIMO}-{OFDM},'' \emph{IEEE Wireless Commun.
  Lett.}, Apr. 2021.

\bibitem{NIPS2011_4209}
S.~Ding, G.~Wahba, and J.~Zhu, ``Learning higher-order graph structure with
  features by structure penalty,'' in \emph{Advances in Neural Information
  Processing Systems 24}.\hskip 1em plus 0.5em minus 0.4em\relax Curran
  Associates, Inc., 2011, pp. 253--261.

\bibitem{8421267}
J.~{Choi}, ``{On} simultaneous multipacket channel estimation and reception in
  random access for {MTC} under frequency-selective fading,'' \emph{IEEE Trans.
  Commun.}, vol.~66, no.~11, pp. 5360--5369, Jul. 2018.

\bibitem{press2007numerical}
\BIBentryALTinterwordspacing
W.~Press, W.~H, S.~Teukolsky, W.~Vetterling, S.~A, and B.~Flannery,
  \emph{Numerical Recipes 3rd Edition: The Art of Scientific Computing}.\hskip
  1em plus 0.5em minus 0.4em\relax Cambridge University Press, 2007. [Online].
  Available: \url{https://books.google.com/books?id=1aAOdzK3FegC}
\BIBentrySTDinterwordspacing

\bibitem{Bertsekas99}
D.~Bertsekas, \emph{Nonlinear Programming}.\hskip 1em plus 0.5em minus
  0.4em\relax Athena Scientific, 1999.

\end{thebibliography}

\end{document}